\documentstyle[12pt,aaspp4]{article}
\lefthead{Hurley-Keller, Mateo, \& Nemec}
\righthead{The Star Formation History of Carina}

\begin{document}

\title{The Star Formation History of the
     Carina Dwarf Galaxy}

\author{Denise Hurley-Keller \& Mateo Mateo}
\affil{Astronomy Department, University of Michigan}
\authoraddr{Ann Arbor, MI 48109}

\author{James Nemec}
\affil{International Statistics \& Research Corporation}
\authoraddr{P.O. Box 496, Brentwood Bay, BC, V8M 1R3}

\begin{abstract}
We have analyzed deep $B$ and $V$ photometry of the Carina dwarf
spheroidal reaching below the old main-sequence turnoff to $V \sim
25$.  Using simulated color-magnitude diagrams to model a range of
star formation scenarios, we have extracted a detailed, global star
formation history.  Carina experienced three significant episodes of
star formation at $\sim 15$ Gyr, 7 Gyr, and 3 Gyr.  Contrary to the
generic picture of galaxy evolution, however, the bulk of star
formation, at least 50\%, occured during the episode 7 Gyr ago, which
may have lasted as long as 2 Gyr.  For unknown reasons, Carina formed
only 10-20\% of its stars at an ancient epoch and then remained
quiescent for more than 4 Gyr.  The remainder ($\sim 30\%$) formed
relatively recently, only 3 Gyr ago. Interest in the local population
of dwarf galaxies has increased lately due to their potential
importance in the understanding of faint galaxy counts.  We surmise
that objects like Carina, which exhibits the most extreme episodic
behavior of any of the dwarf spheroidal companions to the Galaxy, are
capable of contributing to the observed excess of blue galaxies at $B
\sim 24$ only if the star formation occurred instantaneously.

\end{abstract}
\keywords{galaxies: evolution, galaxies: stellar content}

\section{Introduction}
The Carina dwarf galaxy, one of the nine known dwarf spheroidal
companions to the Galaxy, has a surprisingly complex star formation
history.  Detection of carbon stars provided the first suggestion of a
significant intermediate-age population (Cannon, Niss, \&
Norgard-Nielsen 1981), followed by main-sequence photometry which
revealed a young turn-off due to a population perhaps only 6--9 Gyr
old (Mould \& Aaronson 1983; MA hereafter).  MA fit a simulated
luminosity function comprised of 7 Gyr-old stars to their data, and
estimated the contribution of an old population to be relatively small
in comparison.  Saha, Monet, and Seitzer (1986) observed a large
number of RR Lyraes in the central region of the galaxy and
established a lower limit of 2--3\% for the fraction of old stars,
depending on the yield of RR Lyraes. Mighell (1990) estimated the
relative sizes of the populations from the double-peaked color
distribution near the MSTO region to be 85\% for the intermediate-age
burst and 15\% for the old episode.  Using simulated luminosity
functions, Mighell \& Butcher (1992) later fit intermediate-age burst
models to this deep, main-sequence photometry and estimated an upper
limit to the old population of 40\%.  More recently, Smecker-Hane, et
al. (1994; hereafter SHSHL) resolved two separate horizontal branches
(HBs).  Although the separation is not independent of metallicity
effects, it is very suggestive of the multi-episode nature of Carina's
history.

We are left with an estimate between 2\% and 40\% for the fraction of
old stars, and no clear result on the duration of the star formation
episodes.  The chemical evolution of the galaxy is likewise still in
question.  Spectroscopy of 15 giants in Carina (Da Costa 1994)
resulted in an average [Fe/H] of --1.9, with one giant significantly
more metal poor ([Fe/H] = --2.2).  The excellent areal coverage of the
SHSHL data reveal a very thin giant branch.  Given the huge apparent
age spread, a metallicity spread may be necessary to compensate and
produce the observed narrow RGB.

Improved methods of analysis are needed.  Detailed analyses of stellar
luminosity functions have some advantages over more traditional
isochrone fitting.  These methods, however, were designed for coeval
systems.  Because dwarfs exhibit a range of ages and metallicities,
these approaches are problematic at best, and misleading at worst.  A
much better way to unravel complex star formation histories like those
exhibited by dwarf galaxies such as Carina and Leo~I is to use all of
the information embodied by color-magnitude diagrams.  In these cases,
where the galaxies appear to have experienced bursts at intermediate
epochs, a conventional luminosity function would not distinguish
between the old subgiant branch and the young MS stars, for example.
Color information proves essential to resolve multiple components in
these systems.

Simulated color-magnitude diagrams have been used successfully in
studies of bright stars in dwarf irregulars (Tosi et al. 1991; Tolstoy
1996; Gallart et al. 1996a, 1996b, Aparicio et al. 1997a, 1997b), of
LMC field stars (Bertelli et al. 1992; Vallenari et al. 1996), and of
LMC clusters (Vallenari et al. 1994).  We have deep CCD photometry
(limiting magnitude $V = 24.5$) with reasonably small photometric
errors ($\sim 0.02$ at $V=23$) of three fields in the Carina dwarf
galaxy, reaching well below the old main-sequence turnoff.  Because
the photometry reaches below the old MSTO, these data are well-suited
to a detailed extraction of the star formation history by comparing
model color-magnitude diagrams to that of Carina.

Section 2 is a discussion of the observations and reductions of the
Carina data.  Section 3 outlines the analysis applied to these data,
including the development of the pseudo-LF and the generation of
similulated color-magnitude diagrams.  The results, presented in
section 4, are summarized and briefly discussed in the context of
galaxy evolution in section 5.

\section{Observations}
\subsection{Photometry}
The Carina observations were acquired using a $800~\times~800$ TI CCD
at the CTIO 4m telescope during two runs in March of 1989 and 1990.
The observations cover three fields in $B$ and $V$, located as shown
in Figure 1 and labelled F1, F2, and F3.  Also outlined in this figure
is the region (M) observed by Mighell(1990).  The pixel scale for this
instrument was $0.292\arcsec$ per pixel, so that each of our fields is
15.16~arcmin$^{2}$.  The mediocre seeing permitted us to bin the
images $2\times2$, so that the final scale was $0.58\arcsec$ per
pixel.

Field 1 was observed in 1989; fields 2 and 3 were observed in 1990.
The typical exposure time for each image was 500~s, with the total
exposure time in each filter for each field as listed in Table 1.
Table 1 also lists the coordinates of the fields and the number of
images in each filter. Control field observations, taken $1\arcdeg$
south of Carina, are also summarized in Table 1, and were used to
correct the data for contamination due to foreground galactic stars
and background galaxies.

The images were processed and reduced using standard techniques.
After bias subtraction and flatfielding, the images were registered by
calculating frame-to-frame offsets in x and y position using bright
unsaturated stars.  We then coadded all the frames in each filter to
obtain deep combined images in both $B$ and $V$ for each field.  These
combined images were then reduced using DoPHOT (Schechter et
al. 1993).  Athough the conditions were photometric for both runs, the
seeing was worse than $1.2\arcsec$ for some of the exposures which
undoubtedly affected the quality of the combined images.  Also, the
first 100 columns of the images were unusable because of a defective
anti-reflective coating on the chip; thus we excluded these columns
from the reduction.  The number of stars detected in each field is
shown in Table 1.

Each field was calibrated via observations of Graham standards (Graham
1982) which covered an appropriate range in color and airmass.  These
primary standards were used to calibrate the individual B and V frames
for each night.  In general, the fits to the standard star magnitudes
and colors exhibit an rms scatter of $\sigma \leq 0.03$ mag (see Mateo
et al. 1991 for a discussion of the calibration of the 1990 data).
Secondary standards from each field were then used to calibrate the
coadded frames.  Table 2 gives a sample of the $B$ and $V$ photometry
of the Carina stars. The complete version of Table 2, showing
photometry for all the stars detected in all three fields, can be
accessed at ftp:/ra.astro.lsa.umich.edu/pub/get/denise/carina.phot.
Table 3 lists the internal errors as a function of magnitude as
returned by DoPHOT for each field.  Figures 2 and 3 show the
distributions of error vs. magnitude for each of the three fields in
Carina.  The individual frames were also reduced and were searched for
variable stars; those results are discussed in Mateo et al. (1998).

\subsection{Completeness}
Carina is a completely uncrowded field to the limit of our photometry.
The resolved stellar density in the deepest field (field 1) is about 1
resolved star per 52 pixels.  The typical stellar image in that field
covers an area of 28 pixels.  While this does not mean that blending
of stellar images or mismatches of stars between filters does not
occur at all, it does mean that they happen rarely, and thus likely
produce a small effect only on the photometry near the detection limit
(taken to be $3\sigma$). In our analysis, we only consider photometry
roughly one magnitude or more brighter than the detection limit in our
data.

Due to the lack of crowding in the Carina fields, we felt that the
observational effects relevant to the determination of the star
formation history of Carina could be realistically treated without
resorting to the statistical use of artificial stars as described in
Aparicio \& Gallart 1995.  This highly successful but computationally
expensive technique has been applied to several dwarf irregulars in
the Local Group, where the stellar density is much higher and the
contribution of unresolved light much greater.  In our analysis, we
will incoporate completeness effects and photometric errors and
neglect effects due to crowding.

To this end, we conducted a false-star analysis to determine the
completeness as a function of $B-V$ color and $V$ magnitude.  For each
field, 100 $B$ and $V$ image pairs were generated to which 100 false
stars had been added. Thus, the number of false stars added to a
{\it{single}} frame was less than 10\% of the number of real stars
detected, but a total of 30,000 stars were added to all three fields.
The final result is based on rejecting stars added to severely
saturated regions of the images, as well as to the 100 bad columns.
In this way, we established a statistically sound completeness
estimate without significantly altering the degree of crowding.

$V$ magnitudes and $B-V$ colors were assigned to the false stars
beforehand, and the $B$ and $V$ images were generated
{\it{simultaneously}}.  Adding false stars to the $B$ and $V$ images
separately leads to problems with cross terms when calculating the
completeness factor (see Aparicio et al. 1995). The slope of the
luminosity function of artificial stars was fixed to be $\sim0.4$,
roughly that of the data, in order to correctly reproduce the effects
of bin migration (Mateo 1988).  The assigned colors sampled a uniform
distribution across the range covered by the data.  Positions were
also randomly selected from a uniform distribution across the CCD
field.  DoPHOT was then applied to the these frames in the same manner
as to the original data.  Figure 4 shows the differences in input and
output magnitude plotted against the input magnitude for field 1.
This distribution is not symmetric because, while only random error
makes a measured magnitude fainter than an input magnitude, artificial
stars can be added on top of an existing faint, real star, causing
measured magnitudes to be brighter.  While this is a real effect that
will cause the errors to deviate from a gaussian distribution, note
that there are relatively few (less than 10\%) of these stars with
$m_i-m_o$ much greater than zero at magnitudes brighter than $V \sim
24$.

Finally, the completeness is calculated in a simple rectangular grid
in $V$ vs. ($B-V$) by comparing the number of stars added to a grid
element to the number from that grid element which are detected.
Eventually, this completeness factor is used by the program which
generates synthetic color-magnitude diagrams to decide whether a given
star is detected or not.  As we will discuss in section 2.4, we found
the three fields to have no significant differences, and combined
results from them for a quantitative analysis of the global star
formation history.  The combined completeness for the three fields can
be calculated from the results for the individual fields:
\begin{displaymath} 
C(V,B-V)=\frac{C_{1} +  C_{2} + C_{3}}{N_{1} + N_{2} + N_{3}}
\end{displaymath}
where $C$ and $N$ are the completeness and the number of stars added
in a given $V$, ($B-V$) bin for each of the three fields, respectively.
Results of this completeness analysis are shown in Figure 5.  We shall
incorporate these corrections into the model calculations as described
in a later section.

No false star analysis was run on the control field.  The completeness
was assumed to be the same as for the combined Carina images.  This
assumption is inaccurate at faint magnitudes, as the overall depth of
the control field images is less than that of the Carina
images. However, the effect of such an error in the completeness for
this field is diminished by the greater density of Carina stars in the
region of the CMD used in our analysis, and by the fact that only
stars with $V<24.2$ are considered.  Any error is likely to have only
an insignificant effect on the results of our analysis.

\subsection{Comparison with Previous Photometry}
Our photometry shows good agreement with previous photometric studies
of Carina.  The overall calibration of the color-magnitude diagram for
all three fields (Figure 6d) is consistent with the results of MA and
Mighell (1990).  Both put the location of the red clump at $V \sim
20.5$. The average $V$ magnitude of stars in the clump in the CMD in
Figure 6d is 20.46. Because of the greater areal coverage of their
study, SHSHL were able to identify the blue HB at $V$ = 20.65 $\pm
0.05$.  Isolating the sparsely populated extended blue HB in our
diagram, the average $V$ is 20.71, which is consistent with the SHSHL
result.

As can be seen in Figure 1, our fields 1 and 2 overlap with the field
observed by Mighell, and a direct comparison of the photometric
datasets can be made.  We show the results of matching stars from
Mighell observations with those of field 2 (which has the largest
overlapping area) in Figure 7.  Our photometry is systematically
brighter in V by 0.034 $\pm 0.003$ mag.  There were two stars with
differences $>$ 5 magnitude; both were faint stars mismatched with
nearby bright stars.  Most of the other objects with differences $>$ 1
magnitude either have relatively nearby companions or are obviously
mismatched.  Although removal of these objects does not remove the
systematic difference in the photometry, an error of this size in our
photometry, if real, will not affect the final results of our
analysis.  

We constructed a luminosity function from our data, as well as from
those of Mighell (1990) and MA.  No correction was made for
completeness or foreground/background contamination in any of the
datasets for this comparison.  The luminosity functions were
normalized to the number of stars in the $V$ magnitude range 21.2 to
22.5.  Figure 8 shows reasonable agreement between all three studies
until the faintest bins where incompleteness becomes important.

\subsection{Color-Magnitude Diagrams}
Figure 6 shows the color-magnitude diagram for the individual Carina
fields and for all three combined (7309 stars).  Three MSTOs can be
identified in the combined CMD: a ``young'' MSTO at $V \sim 22.3$, an
intermediate-age MSTO at $V \sim 22.7$, and an old MSTO at $V \sim
23.2$.  Given the age spread, the narrowness of the RGB at $B-V \sim
0.7$ is suggestive of a metallicity spread as recognized by SHSHL.  In
particular, note the prominent red clump at $V \sim 20.5$ and the
extended blue HB first noted by SHSHL, but only weakly visible in this
diagram due to the smaller sample of stars.  Also visible in the
diagram are foreground galactic stars with $B-V > 0.6$ and background
galaxies with $V > 23$ and extending to very blue colors.

Does Carina have a blue straggler population?  Many stars in Carina's
CMD obey the strict definition of blue stragglers: stars located along
the main sequence above the old turn-off in an ancient population.
However, we argue that most of these are BS pretenders and that it is
much more likely that Carina has formed stars in the recent past
($\sim 1-3$ Gyr).  In globular clusters, Preston et al. 1994) found
that the ratio of blue stragglers to BHB stars is $\sim
0.6$. Moreover, the luminosity function of blue stragglers in globular
clusters increases from a luminosity cutoff at $M_V \sim 1.9$ towards
the ancient main-sequence turn-off point at $M_V \sim 4$ (Sarajedini
\& Da Costa (1991); Fusi Pecci et al. 1992); for Carina, this cutoff
corresponds to $V \sim 22$.  The blue horizontal branch (BHB) is
comprised of those stars blueward of the instability strip, the blue
edge of which must be near $B-V \sim 0.21$ given the expected
reddening towards Carina.  In the Carina CMD, there are 10 stars in
this region.  If blue stragglers of the kind found in globular
clusters exist in Carina, there should only be roughly six such stars,
and they would be buried among the many stars located along the main
sequence between $V \sim 24-22$.  An old BS population cannot explain
the presence of the pretenders, which number close to 50 and reach to
$V \sim 21.2$; only stars younger than 10 Gyr could populate this
region of the CMD at such high density.

Several factors led us to conclude that any variation in star
formation history between the three fields which we have studied is
insignificant.  First, the three fields lack an obvious qualitative
difference.  Visual inspection shows that the same large-scale
features visible in the combined CMD, such as the intermediate-age
turn-off and the red clump, remain in the CMDs for the individual
fields.  Second, Figure 9 shows that the luminosity functions for the
three fields are similar.  Third, the numbers of stars in large
regions of the CMD which represent different evolutionary stages are
consistent from field to field.  Table 4 lists the numbers of stars in
the ``BS'' region, the BHB, the red clump, and in a region at the
upper MSTO for each field in Carina.  For each region, the numbers are
within roughly $2\sigma$ of the average value for that region.  This
should be contrasted with recent evidence for real spatial variations
in the horizontal branch in some dSph (Da Costa et al. 1996).
Ultimately, any differences between the fields are lost due to the
small number statistics; there simply are not enough stars in any one
field to do the quantitative analysis that is carried out on the
combined CMD (see below).

\section{Analysis}

Interpreting the Carina data involved three basic steps.  First, we
determined the pseudo-luminosity function (pseudo-LF; defined below)
for these data, hoping to avoid the problems of a simple luminosity
function analysis.  We then generated a set of synthetic
color-magnitude diagrams, including photometric errors and
completeness effects, which sampled a parameter space appropriate for
our data.  We determined the pseudo-LF for the model CMDs.  Finally,
we calculated the $\chi^2$ value for the model and data pseudo-LFs and
applied a criteria for selecting the star formation scenarios which
were most likely to lead to the observed Carina CMD.  This method
provides increased detail in the derivation of the SFH of Carina at
the expense of decreased reliablility in the specific results due to
the greater statistical demands placed on the data by this procedure.

\subsection{Determining the Pseudo-LF for Carina}

The analysis of stellar populations in nearby, resolved dwarf
spheroidals has been limited in the past to methods applied to
globular clusters, which dSph were expected to resemble.
Historically, the standard technique has been to combine isochrone
fitting (i.e. Flannery \& Johnson 1982) and an analysis of the
features of the color-magnitude diagram.  Relationships between the
observed properties of the color-magnitude diagram (RGB color at the
level of the HB, for example) and the underlying evolutionary
parameters (metallicity and age, for example) are derived from
galactic globular cluster observations, and observations of globular
cluster systems in nearby galaxies (Sandage 1982; Lee et al. 1990; Da
Costa \& Armandroff 1990, for example).  This approach is designed for
coeval systems, and is problematic for dwarfs which exhibit an
internal range of ages and metallicities.

Detailed analysis of stellar luminosity functions is an improvement at
some level.  Included in the fitting is the relative numbers of stars
at different stages along the isochrone, as well as the shape of the
isochrone.  In addition, uncertainties in the details of stellar
atmosphere models which can affect the colors do not affect the
luminosities strongly (Paczynski 1984; Ratcliff 1987).  Evolutionary
parameters can be derived from the features of the stellar LF.  With
the availability of moderately fast workstations, grids of simulated
LFs for populations with different star formation histories and
chemical evolution can be computed for statistical comparison to the
observations (Mighell \& Butcher 1992; references therein).  This
method is appropriate for populations such as globular clusters which
experience a single episode of star formation, but luminosity
functions for systems with more complex star formation histories are
difficult to interpret.  Intermediate-age populations have been
observed in both the Carina and Leo~I dwarf spheroidal companions to
the Milky Way.  In these cases, where the galaxies appear to have
experienced bursts at intermediate epochs, a conventional luminosity
function would not distinguish between the old subgiant branch and
somewhat younger MS stars.  Color information proves essential to
resolve multiple components in these systems.

We seek to solve this problem by isolating in color the main sequence
turn-off region and the subgiant branch, where evolution slows as
stars first ascend the RGB.  A high degree of resolution in age and
metallicity will be achieved in these two areas where the shape of the
isochrones are sensitive to age and metallicity {\it{and}} the
evolution is relatively slow enough to ensure a statistically
significant number of stars in the studied region. In contrast, the
structure of the unevolved main sequence is dominated by the mass
function, and the RGB by dispersion in color due to the photometry.
In the Hertzsprung gap, the magnitude of stars is very sensitive to
age and metallicity, but the evolution is rapid, so that in our data
there are too few stars in this region for a quantitative analysis.

In a further effort to optimize the sensitivity to dispersion in age
and metallicity, a pseudo-luminosity function is constructed in these
regions.  This function differs from the standard luminosity function
in two ways.  First, if the bins are all of fixed size, the
statistical fluctuations are larger in bins where there are fewer
stars.  Instead, we use bins of variable size which contain a fixed
number of Carina stars. The size of each bin varies and reflects the
stellar surface density in the CMD.  The number of {\it{Carina}} stars
is determined by correcting for the number of contaminants in the bin
using the control field photometry; as the number of control stars in
a given bin varies, so does the size of the bin. This somewhat
complex, iterative procedure ensures a statistically significant
number of stars in each bin.  Second, the slope of the bins is roughly
aligned with the slope of the isochrones in the targeted regions to
prevent the blending of stars from regions of different evolutionary
stages.

We define the pseudo-luminosity function as the ratio of the number in
the bin to the area of the bin, and normalize this to a region which
combines the two brightest bins of the subgiant branch pseudo-LF:
\begin{displaymath} R_{i} = \frac{(N_{i}/A_{i})}{(N_{RGB}/A_{RGB})}
\end{displaymath} where $A_i$ is the area in the indicated bin and $N_i$ is 
the number of stars in that bin.  After some experimentation, 50
stars/bin in the MSTO region and 30 stars/bin in the subgiant branch
region provided sufficient stars per bin to reduce the statistical
fluctuations without smoothing out the important features.  The Carina
CMD with the bins generated in this manner is shown in Figure 10.

The pseudo-LFs for the data, in Figure 11, can be interpreted in terms
of Carina's SFH. The intermediate-age main sequence turn-off appears
as a sudden drop in the MS pseudo-LF at $V \sim 23$.  The old turn-off
is dominated by the intermediate-age main sequence and completeness,
which drops sharply ($\sim 20\%$) between $V = 23$ and $V = 24$.
Keeping this in mind, the old turn-off can be discerned at $V \sim
23.5$ where a less distinct drop occurs.  The subgiant branch is
easier to interpret.  The two obvious peaks correspond to an
enhancement in the number of stars at roughly 3 Gyr and 7 Gyr.  These
populations can also be recognized directly in the varying size of the
bins along the subgiant branch in Figure 10.  The third, sixth, and
seventh bins are noticably smaller than the adjacent bins,
corresponding to the region where 3 Gyr and 7 Gyr isochrones would
join the RGB. The sharp rise at faint magnitudes is where the main
sequence begins to affect the bins.

Since a significant population as young as 3 Gyr is not immediately
obvious in the Carina CMD, we initially considered the possibility
that the brighter peak in the subgiant branch pseudo-LF was the
``Thompson bump''.  A slow-down in the giant branch evolution occurs
when the H-burning shell reaches the discontinuity in chemical
composition left behind by the recession of the convective envelope
(Bergbusch \& VandenBerg (1992) and references therein).  This causes
a peak in the RGB luminosity function, the position and amplitude of
which depend on metallicity.  Given that Carina is extremely
metal-poor ([Fe/H] $\sim$ --2.0 to --2.2 dex), any ``bump'' due to the
old population should be of small amplitude and should be located
roughly two magnitudes brighter than the bright peak in the subgiant
branch pseudo-LF (Fusi Pecci et al. 1990).  Younger populations
produce an even brighter ``bump''.  Thus, neither the old nor the
intermediate-age population can provide an explanation for this
feature.

\subsection{Generating Simulated CMDs}

The Carina CMD simulations use a set of metal-poor isochrones with a
helium fraction Y = 0.235, ranging in age from 2--18 Gyr and in
metallicity from [Fe/H] = --1.45 to --2.2 dex.  The isochrones, kindly
provided to us by Mike Trippico, were generated using a code similar
in physics and techniques to that of VandenBerg (1983).  While
standard techniques are used to convert evolutionary tracks into
isochrones in the fundamental log $T_{eff}$ -- log $L$ plane, the
colors are determined from detailed synthetic spectra which are
computed at intervals along the isochrones (Tripicco, Dorman and Bell,
1992; hereafter TDB).  The evolution extends through the first-ascent
red giant branch (RGB) and terminates with the He-flash.  We do not
use the He-burning tracks also discussed in TDB.  TDB test the models
on M67, a well-studied, old open cluster, and find reasonably good
agreement between the observed RGB and the theoretical RGB which uses
the semi-empirical surface pressure boundary conditions devised by
VandenBerg (1992).

Our program generates synthetic CMDs in the following manner.  A mass
is chosen according to the Salpeter (1955) mass function.
\begin{displaymath} dN = AM^{-x}dM \end{displaymath}
where $A$ is a normalization factor which depends upon the upper and
lower mass limits used in the models.  The limits ($0.7 M_{\odot}$ to
$1.5 M_{\odot}$) bracket the range of masses present in the main
sequence and red giant branch ($V \leq 25$).  The slope of the mass
function, $x$, is 2.5.  We fixed this parameter in order to suppress
the number of models generated and because no strong evidence exists
for varying the slope in old or intermediate-age stellar populations
(Hunter et al. 1997).  Different values of the IMF slope would affect
most strongly the relative strengths of the populations. The models
are normalized to the data using the number of stars in the RGB
brighter than $V = 22$.  This required isolating the red giant branch
from the horizontal branch and obvious contamination.  Combining the
data from all three of the Carina fields yielded 163 stars brighter
than $V = 22$.

We use a Monte Carlo method to randomly select an age according to
either a constant or episodic star formation history.  The metallicity
follows from a linear age-metallicity relationship established by
input parameters.  Given our photometric errors and especially the
small number of giants in our sample, we cannot easily distinguish
between a scenario with no metallicity spread and one with a 0.2 dex
enrichment between the old and intermediate-age populations.  SHSHL
suggested this spread based on the narrowness of their much more
densely populated RGB.  We chose an [Fe/H] = --2.23 for the old
population and [Fe/H] = --2.0 for the intermediate-age and young
populations.  Using a constant [Fe/H] for the duration of star
formation in Carina is unlikely to change the results of our analysis.

Given a mass, age, and metallicity, the photometric properties of the
star are interpolated from the bounding isochrones using equivalent
evolutionary points in a method similar to that described in Bergbusch
\& VandenBerg (1992).  Figure 12 shows an isochrone interpolated from
bounding isochrones and one derived from the $log L - log T_{eff}$
plane.  At the points of least agreement, the color error is $\leq
.02$ and the error in $V$ is $\leq 0.02$.

The program also incorporates observational effects into the CMD.  We
supply the appropriate magnitude and color errors as a function of
magnitude, estimated from the internal errors returned by DoPHOT (see
Table 3).  The program then randomly assigns error in both magnitude
and color independently, assuming a Gaussian distribution.  The
completeness as a function of both magnitude and color is determined
from the false star analysis.  A grid of completeness values is
supplied to the program.  The star is either kept or discarded based
on the completeness interplolated from the grid given its color and
magnitude.  Rather than correcting the model color-magnitude diagram
for contamination due to background galaxies and foreground stars, we
correct the pseudo-luminosity function directly later in the analysis.

The full star formation history of a galaxy is a potentially complex
function of time and location. To make the problem manageable, we were
forced to limit the range of parameter space which our models covered
based on previous results and the suggestions of our own data.  The
color-magnitude diagram of Carina suggests episodes of strong star
formation, bracketed by quiescent periods.  Main-sequence turn-offs
reveal convincingly the presence of old ($\sim 15 $Gyr) and
intermediate-age ($\sim 7$ Gyr) populations. The inability of any
two-episode models to reproduce the brighter peak in Carina's subgiant
branch pseudo-LF led us to the suspect the existence of a ``young''
($\sim 3$ Gyr) population.  As a result, we concluded that three
episodes of star formation were responsible for the bulk of Carina's
stars.  Figure 13 illustrates the parameterization of Carina's star
formation history.  We focused on extracting the age and SFR
information: $t_{i}$ designates the age center of the SF episode,
$\Delta_{i}$ the duration of the episode, and $S_{i}$ the relative
strength of the episode.

Photometric errors and age resolution conspire to limit our
understanding of the old population.  The suspected strength of this
population ($\leq 30\%$) means there would be a relatively small
number of old stars which, depending on the episode duration, could be
lost in the statistical noise of the main sequence at that magnitude.
In addition, the photometric errors cause a 0.5 Gyr-wide episode to
appear 2 Gyr-wide in in the CMD. Thus, constraining the episode center
and duration is difficult for this population.  The pseudo-LF will be
sensitive to the population strength, however, as it is normalized to
a region on the RGB which will contain old stars. Thus $t_{3}$ and
$\Delta_{3}$ are fixed at 15 Gyr and 1 Gyr, respectively, and $S_3$ is
allowed to vary between 10\%, 20\%, and 30\%.  While we can still
estimate then the contribution of this population to Carina, we cannot
precisely determine the implied star formation rate because we can
only place an upper limit on the duration of the episode.

For the population near 3 Gyr, the fine age resolution helps to limit
the parameter space that needs to be covered.  Figure 14 shows the
subgiant branch pseudo-LF for Carina and three models which have
strong 3 Gyr populations.  Shifting the center of the episode by 1 Gyr
creates a clear effect in the subgiant branch pseudo-LF.  As a result,
ages of 4 or 2 Gyr are less likely to produce a good fit to the data.
3 Gyr is a reasonable choice for $t_1$.  The episode duration,
$\Delta_1$, and strength, $S_1$, remain free parameters.

We expect to extract the most information about the intermediate-age
($\sim 7$ Gyr) episode.  Previous research as well as the CMD itself
indicate that this is the largest population in Carina, so that there
will be a significant number of stars to populate the bins.  In
addition, a reasonable degree of age resolution is still possible at
this magnitude given the photometric errors.  Therefore, $t_2$,
$\Delta_2$, and $S_2$ are free parameters, covering values appropriate
given the appearance of the color-magnitude diagram.  Tables 5 and 6
show the range of parameters covered by the 180 models generated for
this analysis.

Photometry of a large number of stars is needed to damp the
statistical poisson noise.  Figure 15a shows the pseudo-LFs of five
models with the same star formation history with roughly the same
number of stars as the Carina data.  The large amount of variation
between these models is apparent.  The statistical variations between
them limit how well we can distinguish between distinct models.  We
cannot easily determine what effects are due to statistical variation
and what to the difference in star formation history.  Figure 15b
shows the pseudo-LFs for five models with the same star formation
history, but each with ten times as many stars as the Carina data.  As
expected, the variations are much smaller.  Thus, using a model ten
times the size of our data set, we are reducing the relative size of
the errors on the model pseudo-LFs, and improving the sensitivity of
our method.  Each of the model CMDs in Table 6 has 1630 stars in the
RGB above $V = 22$.

Finally, the model RGBs appeared about 0.04 magnitudes bluer than the
data.  The exact cause of the shift is unknown, although several
possibilities exist.  The Trippico et al. isochrones are derived from
Vandenberg isochrones, which are historically believed to be too blue
by roughly the observed amount (Vandenberg 1983).  Another possible
source of discrepancy is the calibration of our photometry, which was
found to be systematically brighter by about 0.03 magnitudes than
Mighell's $V$ photometry.  If Mighell's magnitudes are correct, then
assuming our $B$ magnitudes are correct, our $B-V$ colors would be too
red by 0.03.  Lastly, our chosen values for $z$ in the synthetic CMDs
could be too metal poor to accurately describe the data.  Whatever the
source of the discrepancy, the bins as applied to the models are
shifted in color 0.04 magnitudes to the blue.  Without the shift, fits
for all the models are uniformly worse.  Within the framework of our
approach, one could consider the color shift to be an additional
global parameter that we have to determine.

\subsection{Comparing the Data and the Models}

The $\chi^2$ statistic can be used to compare the pseudo-LFs of the
models and the data.  \begin{displaymath} \chi^{2} = \frac{(R_{data} -
R_{model})^2} {\sigma_{data}^2 + \sigma_{model}^2} \end{displaymath}
where $R_{data}$ is the ratio measured from the data, $R_{model}$ is
that measured from the model, $\sigma_{data}$ is the error in
$R_{data}$, and $\sigma_{model}$ is the error in $R_{data}$.  Both
$\sigma_{data}$ and $\sigma_{model}$ are calculated assuming Poisson
counting errors for the pseudo-LF bins.

We tested the ability of our method to correctly identify the
underlying star formation scenario by generating ten models with the
same input SFH parameters and roughly the same number of RGB stars as
the Carina data.  Each of these ten sets of data were treated as
real data; i.e.  bins were determined so that the same number of stars
were in each bin, and these bins were then applied to the set of 
candidate models described in Tables 5 and 6.  

We then arbitrarily accepted models for which the probability of
measuring a value of $\chi^2$ greater than the $\chi^2$ calculated for
the model was better than 70\% for both the subgiant branch fit and
the main sequence fit.  In this case, only three of the trials yielded
an acceptable solution.  Two of the three yielded unique solutions
which were not the input star formation history, but were only
different in the duration of the young burst (0.5 Gyr instead of the
input 1 Gyr) and the strength of the intermediate-age and young bursts
(50\% and 40\% respectively as opposed to the input 60\% and 30\%).
The third produced two solutions, one of which was the input star
formation history, and one which differed only in that the
intermediate-age burst was off-set by 0.5 Gyr. We therefore estimate
that we can, in principle, determine the episode strength to within
10\% and the episode center and duration within 0.5 Gyr, insofar as
the true model is contained within the set of models which we
generated.  Our preliminary analysis has already led us to a
reasonable range of models.

It is not entirely suprising that some of the 10 trial models were not
matched by any of the 180 candidate models, despite the trial models
being drawn from the candidate models.  We are placing great demands
on a relatively small number of stars in our dataset.  Lowering the
criteria would guarantee a match each time, but would increase the
number of false matches.  An acceptable fit can occur for two reasons
- the data were truly produced by one of the input star formation
history covered by the candidate models, or the data were produced by
some other star formation history and an acceptable fit resulted
because of stochastic effects.  The cutoff criteria were set at a high
percentage to eliminate as many "false" matches as possible.  If the
criteria resulted in no correct matches for our Carina data, we would
then be forced to lower the criteria, and eventually at some point to
consider a new set of candidate models.  This last possibility would
probably not be necessary given that we chose the set of models based
on a preliminary analysis of the Carina CMD.

\section{Results}

Having established that the selection criteria in the previous section
worked reasonably well, we applied the same criteria to the 180 large
models generated for comparison with our data.  The best fit models,
those with the probability of $chi^2$ for both the main sequence and
subgiant branch fits better than 70\%, are shown in Table 7. The main
sequence and subgiant branch $\chi^2$ probabilities for each model
were averaged, and this value was used to rank the models in this
subset.  In addition, models with the probability of $chi^2$ for both
the main sequence and subgiant branch fits better than {\it{50\%}} are
also listed.  The individual models are not as important as the
general properties of models which passed the selection criteria.
These characteristics are summarized in Table 8, and examined in
detail below.

In order to fully demonstrate the goodness of the best fits, we show
in Figure 16 pseudo-LFs for three models which did not pass the cut.
Discrepancy exists both in the normalization and in the shape of the
pseudo-LFs.  Recall that due to the way in which we define the
pseudo-LFs, the normalization is {\it{not}} a free parameter; the
pseudo-LFs cannot be shifted vertically at all in Figure 16.
Qualitative discrepancy is found in the pseudo-LFs of all three
models.  Some of the effects are systematic in nature, i.e. related to
variations in the parameters, and others are statistical, i.e. related
to the stochastic noise in the models.  The latter should be
relatively small since the models are 10 times as large as the data.

In contrast, the three best fitting models are compared in Figure 17,
which shows Carina's CMD as well as those of the three best fit
models.  Note that the three CMDs resemble each other and the data
with the exception of somewhat too many stars in the Hertzprung gap at
3 and 7 Gyr.  This could either represent an overestimate of the size
of those populations in our models, or reflect a problem with the
physics of the stellar models during that phase.  Similarly, the main
sequence and subgiant branch fits (Figure 18) are very good.  Not only
the overall shape but also the normalization is correct.  Because
the models are 10 times as large as the dataset, the models are less
noisy at the faint end in the MS pseudo-LF than the data, and are thus
smooth where the data are spiky.  The mean trends, however, are the
same in these models as in the data.

Some properties of the successful models are compared in Table 8.  An
important episode of star formation 7 Gyr ago is one common property
of all of the models listed in table 7.  Table 8 shows that 10 of the 12
models had $t_2$ = 7 Gyr and the remaining two had $t_2$ = 6.5 Gyr.
However, the intermediate-age peak in the subgiant branch pseudo-LF
for the models with $t_2$ = 6.5 Gyr is consistently too bright (Figure
19).  No model survived which had $t_2$ = 8 Gyr.  Thus $t_2$ is
well-constrained to a narrow range around 7 Gyr.  Table 8 also shows
that no value of $\Delta_2$ can be conclusively ruled out.  Given
these results, we were concerned that our choices for $\Delta_2$ may
not have bracketed the true value.  As a test, we generated a model
with the same parameters as one of the three best fit models, but
$\Delta_2$ = 3 Gyr.  It is apparent in Figure 20 that the resulting
intermediate-age peak is now too broad.  The photometric errors make
an episode shorter than 0.5 Gyr impossible to resolve.

The intermediate-age episode is confirmed as the dominant epoch of
star formation in Carina.  Only models with the intermediate-age
episode stronger than the young episode by at least 20\% survived the
selection criteria.  Although exact age information about the old
population was out of reach, we were successful in constraining the
strength of the old population, $S_3$, which was expected to mainly
affect the normalization of the main sequence pseudo-LF.  Table 8
shows that models with $S_3$ = 0.2 were far more likely to produce a
good fit than those with $S_3$ = 0.1 or 0.3.

Detailed information about the 3 Gyr episode of star formation was
elusive, as expected.  The age of the episode is constrained by the
magnitude of the corresponding subgiant branch peak.  But models with
an episode lasting 0.5 Gyr were just as likely to produce a reasonable
fit as those with star formation lasting 1 Gyr.  It is possible that
either our choices for $\Delta_1$ were not near to the true value, or
our method was somehow not sensitive to changes in this parameter.  As
this youngest population accounts for 30\% or less of the total, and
because the mass function and evolution conspire to depopulate this
region, there was a relatively small number of stars with which to
work. Finally, we point out that there is already evidence of an even
younger population ($\le 3$ Gyr) in Carina, as shown in Smecker-Hane
et al. 1996.  However, we have too few stars to constrain such a weak,
young population.

\section{Conclusion}

Carina experienced an episode of star formation 7 Gyr ago which lasted
no more than 2 Gyr and which was responsible for at least 50\% of its
stars.  The old population ($12-15$ Gyr) may amount to $10\%-20\%$ of
Carina. The bulk of the remainder (($20\%-30\%$) is relatively young,
between 2.5 and 3.5 Gyr old.  While the details of one galaxy's SF
history may seem inconsequential, taken in context and as a member of
the Local Group dwarf population, the details are relevant to deeper,
unresolved issues in cosmology and galaxy evolution.

In the case of Carina, all studies confirm that at least one
relatively long pause in star formation lasting $\sim 4$ Gyr occurred.
The mechanism by which this kind of low mass system could experience
such a pause and then another strong burst of star formation is not
understood.  Current ISM simulations can produce such a large gap in
an isolated dwarf only by ejecting the gas out to $\sim 20$ kpc (Babul
1996).  A dwarf such as Carina, residing in the potential well of the
Galaxy, should lose that gas, preventing any further episodes of star
formation.

Recent Hubble Space Telescope observations of Carina have been studied
by Mighell (1997). He claims to detect a significant number of stars
in the region of the gap in star formation lasting from roughly 8 to
12 Gyr ago.  This is interpreted as indicative of significant star
formation which began in the central region of Carina, and propagated
to the outer regions.  However, the number of stars in the same
section of our CMD, derived from fields several arcminutes from the
center of Carina and overlapping the Mighell (1990) field, is
consistent with the number in the HST CMD.  Whether or not the number
of stars in this area implies significant on-going star formation,
there does not seem to be evidence of a difference between the central
region and regions further out, such as that seen in the
recently-discovered Antlia dwarf galaxy (Aparicio et al 1997c).

Assigning ages to individual faint stars based on the isochrones which
constrain them in color and magnitude must be handled with care.
Firstly, at faint magnitudes, stars have a larger photometric error
and thus a larger implied age range.  Age determinations for specific
stars are therefore subject to greater statistical uncertainty.
Secondly, the lifetime of stars crossing the Hertzprung gap increases
as stars become less massive.  This must be accounted for when
predicting the relative contributions of different ages. In light of
these considerations, the results of the WFPC2 study may not differ
from earlier ground based results.

Evidence of intermediate-age stars in the halo leads to the question
of whether dwarf spheroidals shredded by the Galaxy could be a
significant source of the halo population (Preston et al. 1994; Mateo
1996).  The Sagittarius dSph is currently being ripped apart by the
Milky Way, proving that this type of interaction between a large
galaxy and its tiny neighbors does occur (Ibata, Gilmore, \& Irwin
1994; Mateo et al. 1995).  In addition, the dSph, especially Carina
and Leo I, contain significant numbers of intermediate-age stars.
Mateo (1996) shows that the fraction of relatively young and
intermediate-age stars in the entire population of dwarf spheroidals
is not inconsistent with the fraction in the Galactic halo.

The detailed interpretation of deep galaxy counts and redshift surveys
depends on the SFH of the dwarfs which comprise the excess of faint
galaxies.  For example, `flashing' dwarfs - i.e. bursting quickly and
intensely - contribute to the deep counts to a dramatically different
extent than do dwarfs that experience more drawn-out bursts that
slowly turn on and off (Campos 1997).  The prevalence of bursting
behavior in the SFH of Local Group dwarfs may statistically constrain
the degree to which galaxies similar to these local dwarfs could be
responsible for the faint excess.  

We can place an upper limit on the star formation rate (SFR) of Carina
7 Gyr ago by assuming that the episode of star formation took place
instantaneously.  By choosing a reasonable model, we can easily
calculate the total number of stars formed during this episode in the
volume of the galaxy covered by our three fields, taking into account
the incompleteness.  We can use this number to normalize the IMF, and
then integrate to get the total mass formed during this episode.  The
SFR is then given by: 
\begin{displaymath} SFR =
\frac{M_{total}}{\Delta t}*\frac{1}{\gamma   }  M_{\odot  }  yr^{-1}
\end{displaymath} 
where $\Delta t$ is the duration of the episode in
years and $\gamma$ is the fraction of the galaxy's total surface
brightness in the three fields.  If the episode lasted 10 Myr, the
implied SFR for Carina at that time would be roughly 0.8 $M_{\odot  } 
yr^{-1}$.  If the episode lasted 2 Gyr, the SFR is down by more than a
factor of 100 to 0.004 $M_{\odot  } yr^{-1}$.  Either SFR is
comparable to the instantaneous SFR in even such luminous objects 
as blue compact dwarfs (Fanelli et al. 1988).

Could Carina-type galaxies contribute to the excess of faint galaxies
at around $B \sim 24$?  These galaxies have been shown to have
redshifts between 0.3 and 0.7 (Glazebrook et al. 1995).  Assuming
$H_{\circ}$ = 50 km/s/Mpc, and $\Omega$ = 1, 7 Gyr corresponds to $z
\sim 0.5$, which places the episode of star formation in the
appropriate epoch.  If the burst were essentially instantaneous, the
implied total luminosity of Carina at 7 Gyr would be $\sim 2 \times 10^8
L_\odot$.  Carina would be visible to $z \sim 0.3$ in a sample limited
to $V \sim 25$, putting it at the near edge of objects that could
contribute to these counts.  Galaxies 3-10 times more luminous would
fall in the range $z \sim 0.5$ to 1.0.  Although the local dSph cover
a range up to 50 times the luminosity of Carina, Carina is the most
extreme example of this type of episodic behavior. Further, if the
episodes of star formation are extended in time, then these dwarfs
would only be visible locally.  Thus, despite concerted efforts at
unravelling the SFH of these galaxies, their role in faint galaxy
counts problem remains unresolved.

\section{Acknowledgements}
The authors would like to thank Antonio Aparicio for his thoughtful
comments regarding this paper.  DHK thanks Philipe Fischer and Andrew
Layden for their helpful comments, as well.  This research was 
funded in part by NSF.

\newpage
\figcaption {The Carina dwarf galaxy.  Outlined are the three fields
discussed in this paper, labelled F1, F2, and F3.  Each field is 3.89
$\arcmin$ across. Also, the field observed by Mighell (1990) is
labelled M. The orientation is North to the top and West to the
right.  \label{Figure 1}}

\figcaption {Error in $V$ magnitude vs. $V$ magnitude.  \label{Figure 2}}

\figcaption {Error in $B-V$ color vs. $V$ magnitude.  \label{Figure 3}}

\figcaption {Results of the false star analysis which shows the
difference between the input magnitude, $m_i$, and the measured
magnitude, $m_o$, vs. $m_i$ for field 1.
\label{Figure 4}}

\figcaption {Combined completeness factors as a function of $V$
magnitude and $B-V$ color for the three Carina fields.
\label{Figure 5}}

\figcaption {Color-magnitude diagrams for the three fields observed in
Carina.  (a) Field 1, (b) Field 2, (c) Field 3, (d) All three fields
combined.  The solid line represents the limiting magnitude of the
shallowest field, field 3.  
\label{Figure 6}}

\figcaption {Comparison between Mighell's (1990) $V$ photometry and
ours.  $V$ represents the magnitudes presented in this paper, $V_{M}$
those of Mighell (1990).  Our results are systematically 0.034
magnitudes brighter than those of Mighell (1990).  This discrepancy,
if real, is unlikely to affect the results of this study.
\label{Figure 7}}

\figcaption {A comparison of luminosity functions.  The short dashed
line is derived from Mighell's data, the long dashed line from MA
(1983), and the solid line from ours.  No corrections have been made
for completeness or contamination in any of the datasets.  The
observations were normalized to the number of stars in the six
brightest bins.
\label{Figure 8}}

\figcaption {The luminosity functions for the three fields in Carina.
No correction has been made for foreground or background contamination
in these luminosity functions.
\label{Figure 9}}

\figcaption {Pseudo-LF bins in the main sequence and subgiant branch
regions of the CMD.  Each bin on the main sequence contains 50 stars,
and each bin on the subgiant branch contains 30.
\label{Figure 10}}

\figcaption{ Pseudo-LFs for the Carina data, R, vs. the V magnitude 
of the bin center.  Shown are the main-sequence pseudo-LF (bottom) and
the subgiant branch pseudo-LF (top).
\label{Figure 11}}

\figcaption{Comparison of an interpolated isochrone with a model
isochrone provided by M. Trippico.  Solid line: model isochrone with
age = 4 Gyr and [Fe/H] = $-2.23$.  Dashed line: isochrone with the
same age and metallicity interpolated by the program which generates
synthetic CMDs.
\label{Figure 12}}

\figcaption{The parameterization of Carina's star formation history.
We assumed three major episodes of star formation.  $t_i$ is the age
of the episode in Gyr, $\Delta_i$ is the duration of the episode in
Gyr, and $S_i$ is the strength of the episode in a fraction of the total
population.  For reasons explained in the text, $t_1, t_3$, and
$\Delta_3$ are fixed at 3 Gyr, 15 Gyr, and 1 Gyr, respectively.
\label{Figure 13}}

\figcaption{The subgiant branch pseudo-LFs for four models which
differ only in the value of $t_1$.  $t_2$ is 7 Gyr, $\Delta_1$ and
$\Delta_2$ are 1 Gyr, and $S_1, S_2$, and $S_3$ are 0.3, 0.5, and 0.2,
respectively.  Because isochrones are relatively widely separated at
magnitude $V \sim 22$, it is possible to rule out an age of $\sim 4$ Gyr
for the youngest episode of star formation based on the subgiant
branch pseudo-LF.
\label{Figure 14}}

\figcaption{Stocastic variation between models with the same star
formation history.  Panel (a) shows the main sequence pseudo-LFs of
five models with the same input parameters, but different random
number seeds.  These models have the same number of stars in the RGB
region as the Carina data.  Panel (b) shows the main sequence
pseudo-LFs of five models with the same input parameters, but ten
times as many stars in the RGB region.
\label{Figure 15}}

\figcaption{Three models which fit the data poorly.  Panel (a) shows
the main sequence pseudo-LF and panel (b) shows the subgiant branch
pseudo-LF.  Solid: The Carina data.  Dotted: A model with $\Delta_1$ =
0.5 Gyr, $t_2$ = 8 Gyr, $\Delta_2$ = 0.5 Gyr, and $S_1$, $S_2$, and
$S_3$ are 0.2, 0.7, and 0.1, respectively.  Short dash: A model with
$\Delta_1$ = 1.0 Gyr, $t_2$ = 6.5 Gyr, $\Delta_2$ = 1.0 Gyr, and
$S_1$, $S_2$, and $S_3$ are 0.2, 0.5, and 0.3, respectively.  Long
dash: A model with $\Delta_1$ = 1.0 Gyr, $t_2$ = 7 Gyr, $\Delta_2$ =
0.5 Gyr, and $S_1$, $S_2$, and $S_3$ are 0.1, 0.7, and 0.2,
respectively.
\label{figure 16}}

\figcaption{The color-magnitude diagrams of three models which have
the best overall fit to the data.  Note that control field stars have
been added to these model diagrams.  The Carina CMD has three times
the contamination shown in the models. Panel (a) shows the combined
CMD for Carina.  The star formation histories for the model CMDs are
shown in the remaining panels: (b) A model with $\Delta_1$ = 1.0 Gyr,
$t_2$ = 7 Gyr, $\Delta_2$ = 2.0 Gyr, and $S_1$, $S_2$, and $S_3$ are
0.3, 0.5, and 0.2, respectively; (c) A model with $\Delta_1$ = 0.5
Gyr, $t_2$ = 7.0 Gyr, $\Delta_2$ = 2.0 Gyr, and $S_1$, $S_2$, and
$S_3$ are 0.3, 0.5, and 0.2, respectively; (d) A model with $\Delta_1$
= 1.0 Gyr, $t_2$ = 7 Gyr, $\Delta_2$ = 1.0 Gyr, and $S_1$, $S_2$, and
$S_3$ are 0.3, 0.5, and 0.2, respectively.
\label{figure 17}}

\figcaption{The pseudo-LFs of the three models which have the best
overall fit to the data. Panel (a) shows the main sequence pseudo-LFs,
and panel (b) the subgiant branch pseudo-LFs.  The sold line in each
repesents the pseudo-LFs derived from the Carina data. The star
formation histories for the models are explained in the caption to
figure 16.
\label{Figure 18}}

\figcaption{Panel (a) shows the subgiant branch pseudo-LFs of the
Carina data and three models with $t_2$ = 6.5 Gyr.  These models
consistently produce an intermediate-age peak ($V \sim 22.5$) which is
slightly too bright.  Panel (b) shows the subgiant branch pseudo-LFs
of the Carina data and three models with $t_2$ = 8 Gyr.  These models
consistently produce an intermediate-age peak which is slightly too
faint.
\label{Figure 19}}

\figcaption{Subgiant branch pseudo-LFs of the Carina data and a model
with $\Delta_2$ = 3 Gyr.  This model produced an intermediate-age peak
($V \sim 22.5$) which is clearly too broad.
\label{Figure 20}}

\plotone{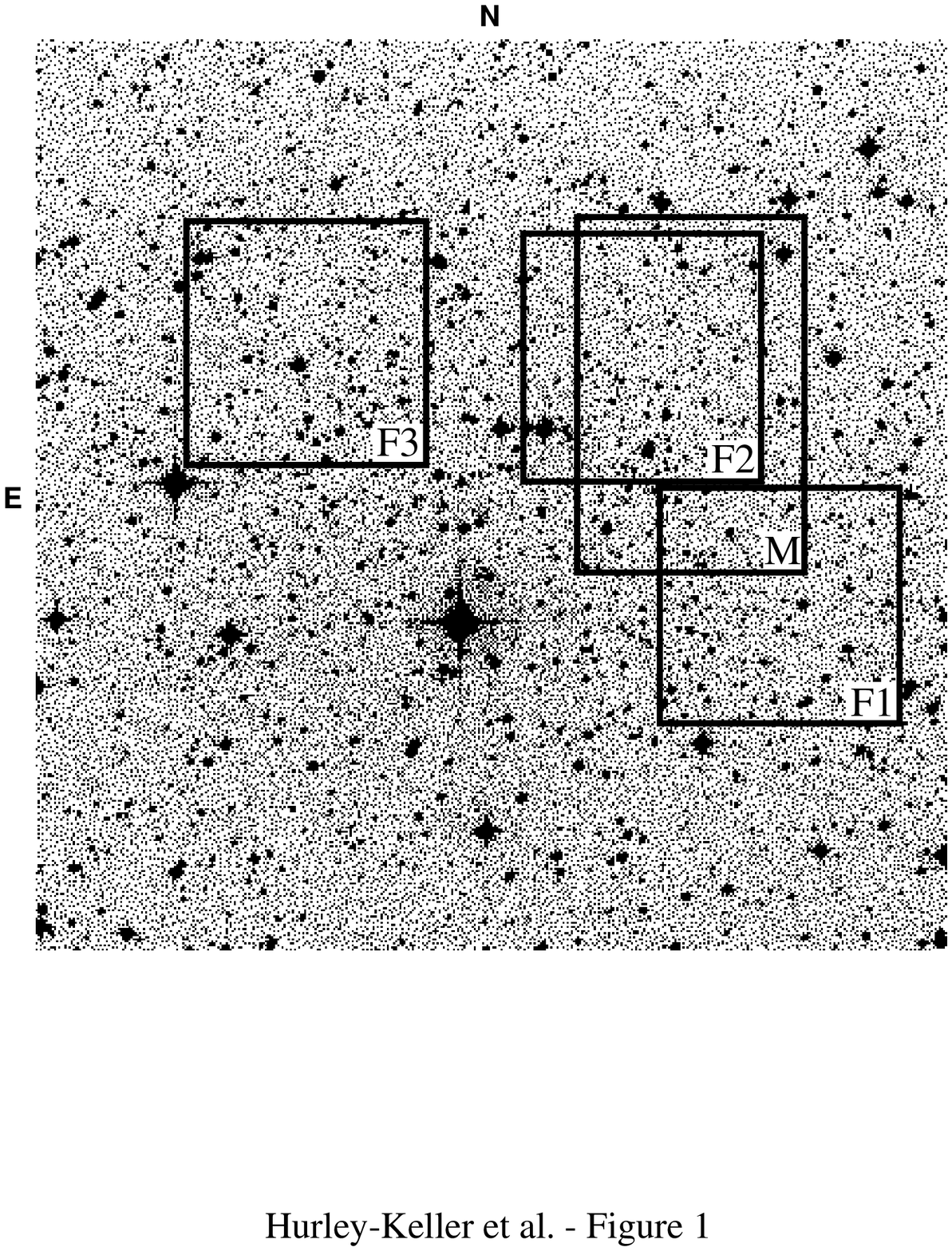}
\newpage
\plotone{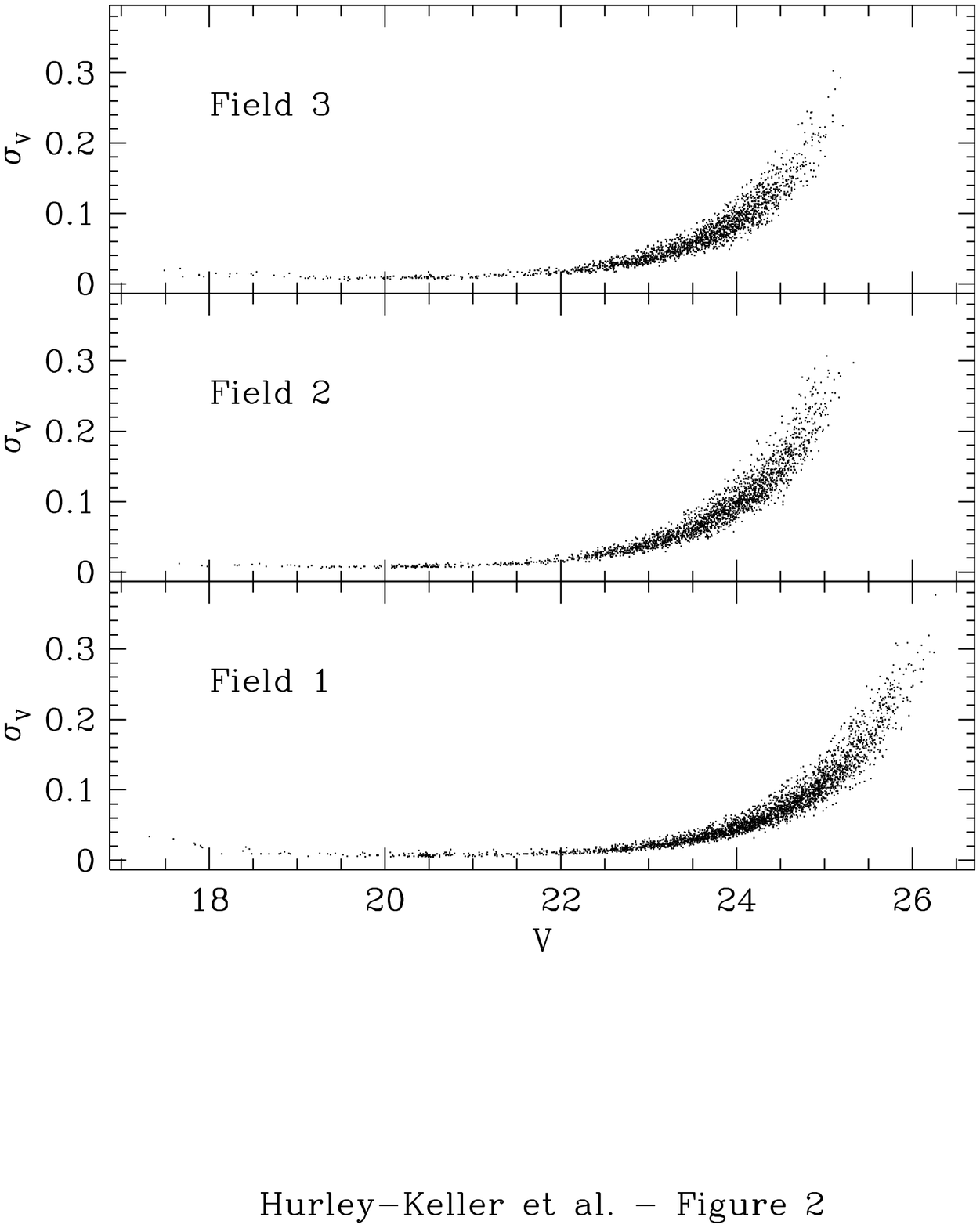}
\newpage
\plotone{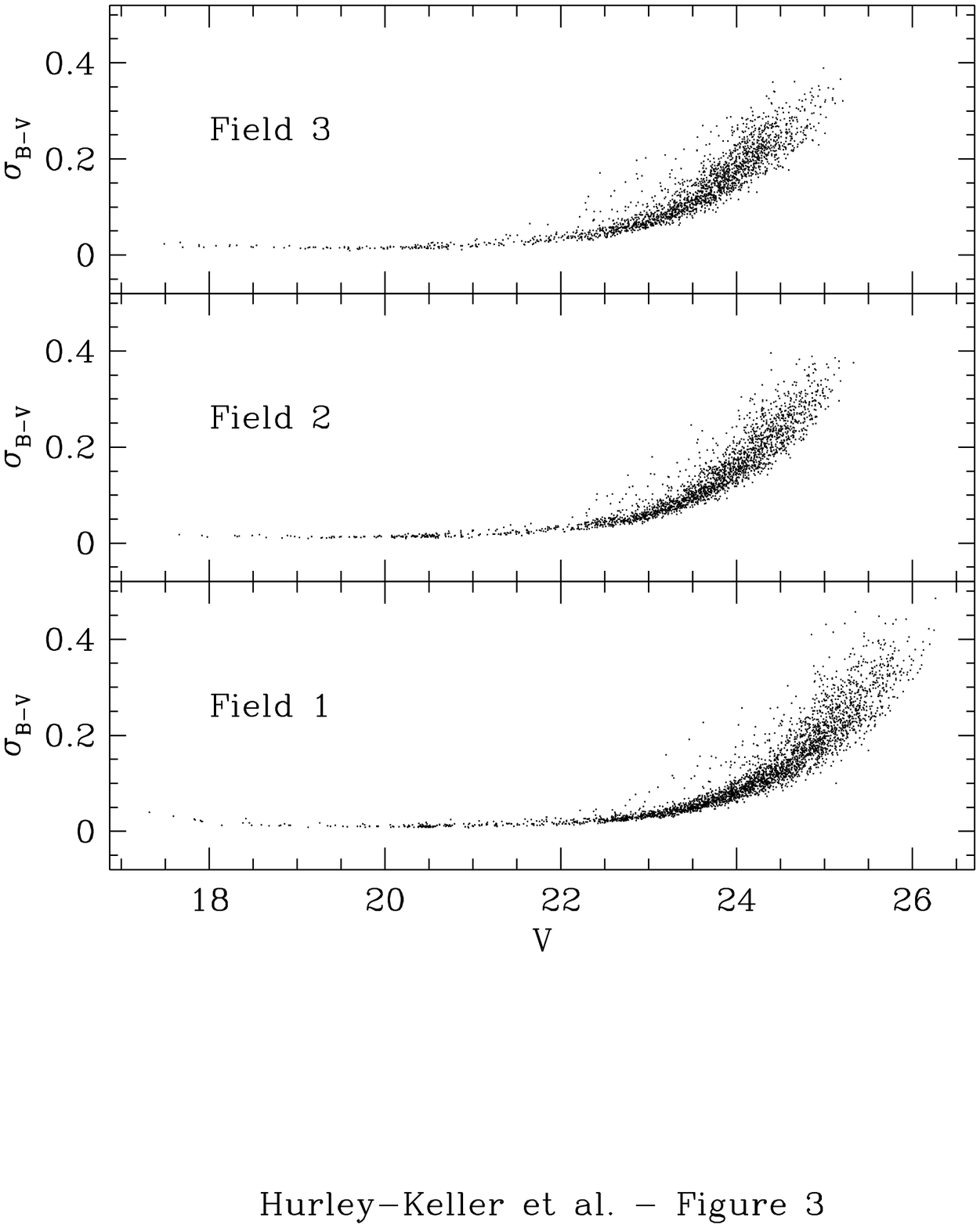}
\newpage
\plotone{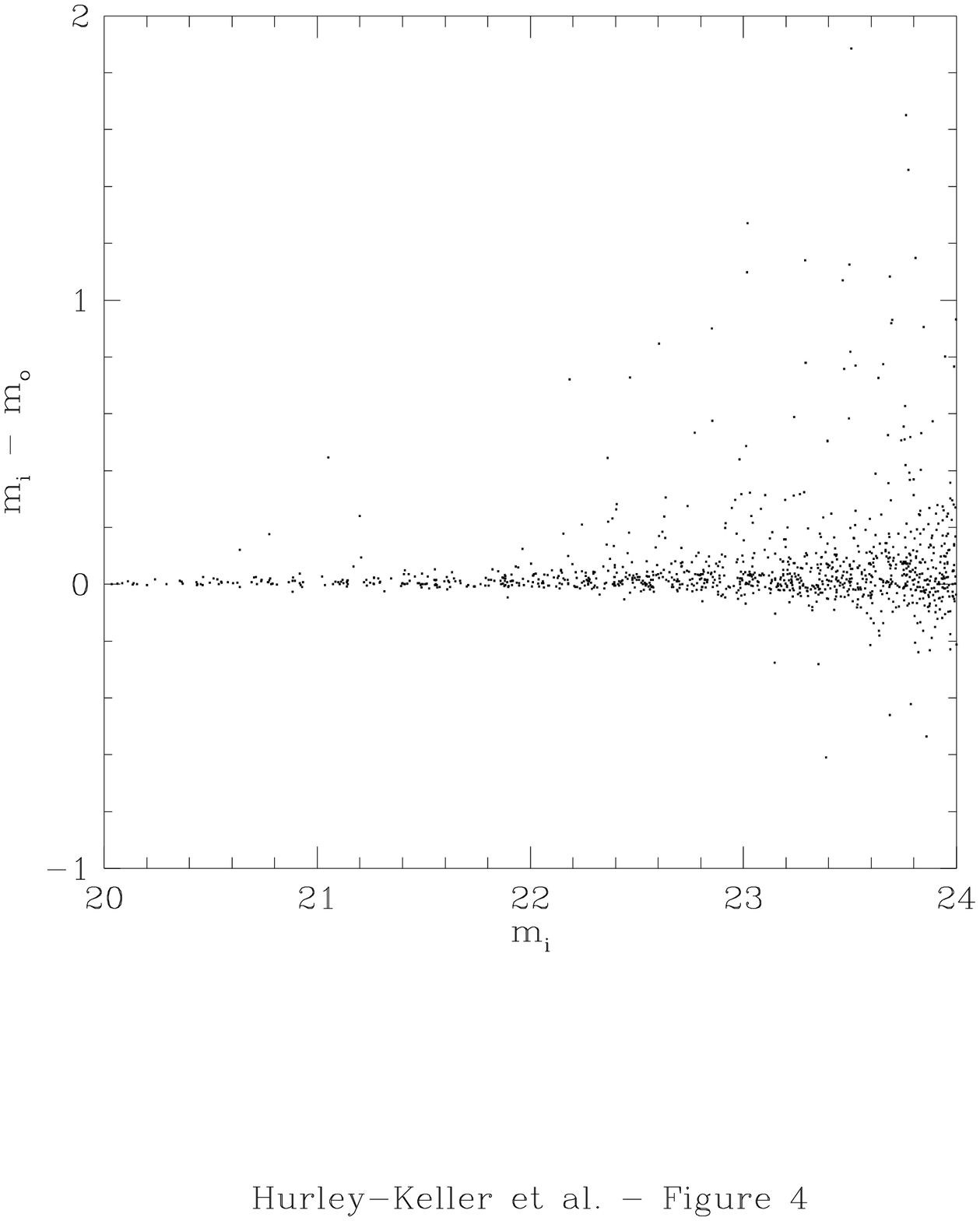}
\newpage
\plotone{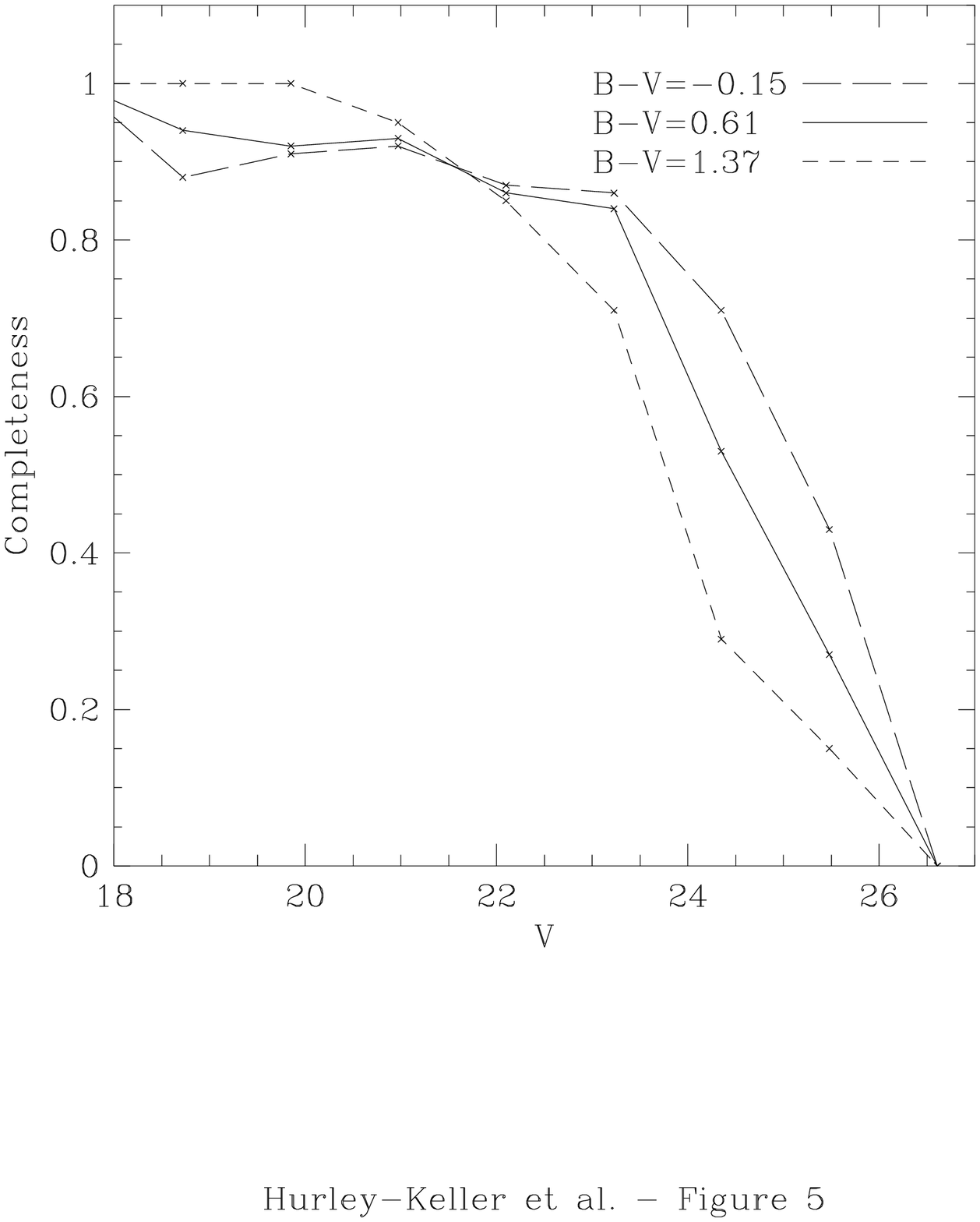}
\newpage
\plotone{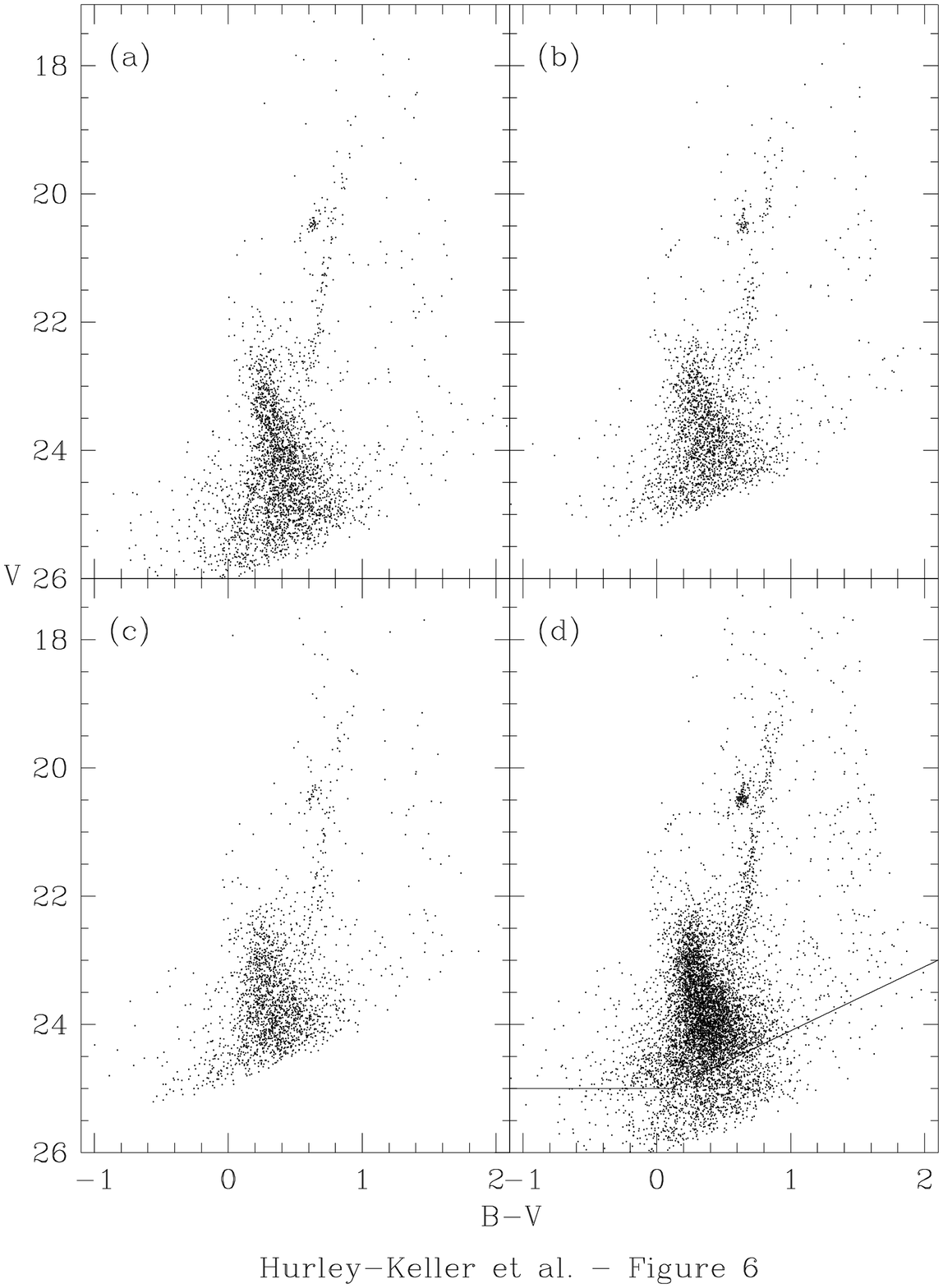}
\newpage
\plotone{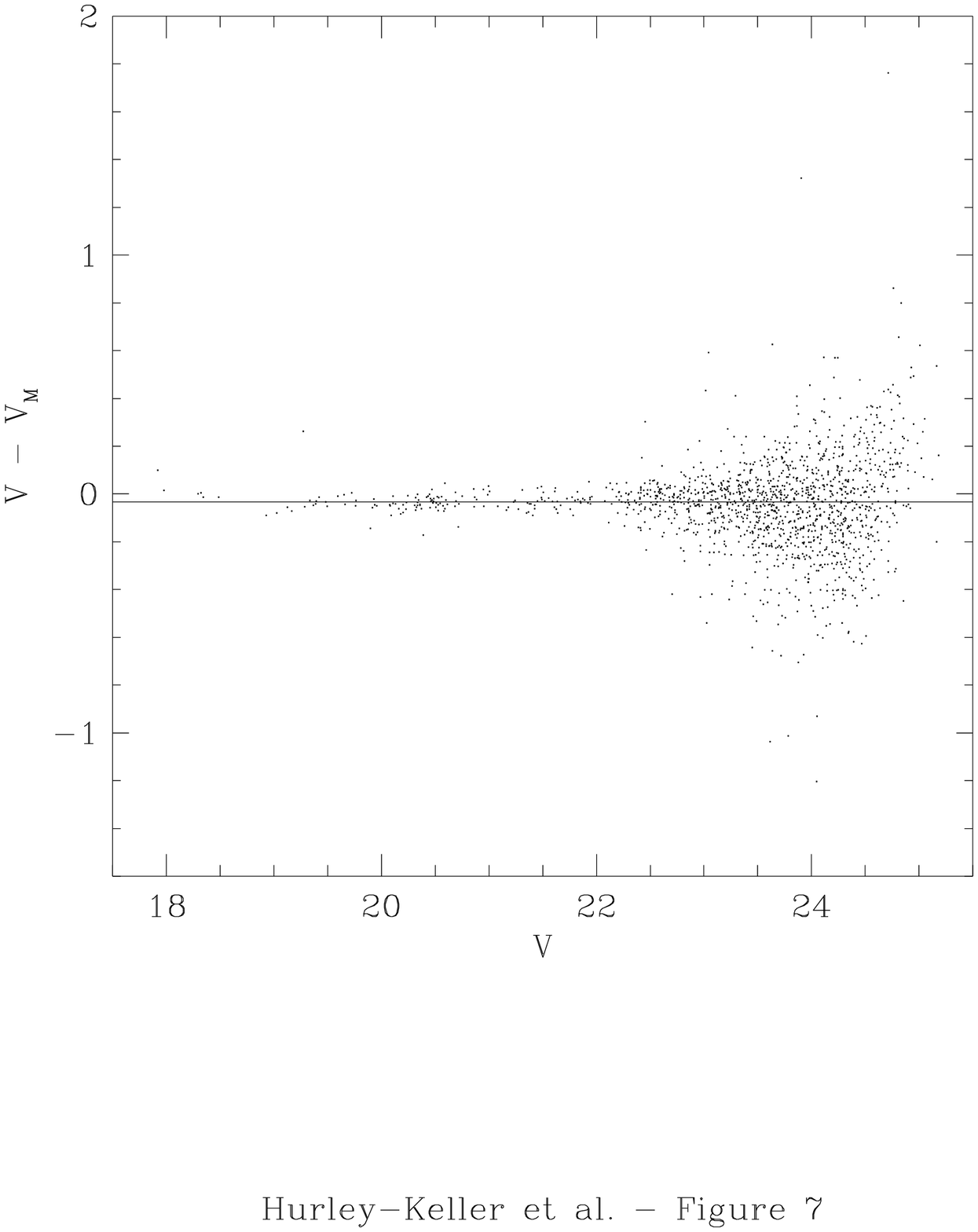}
\newpage
\plotone{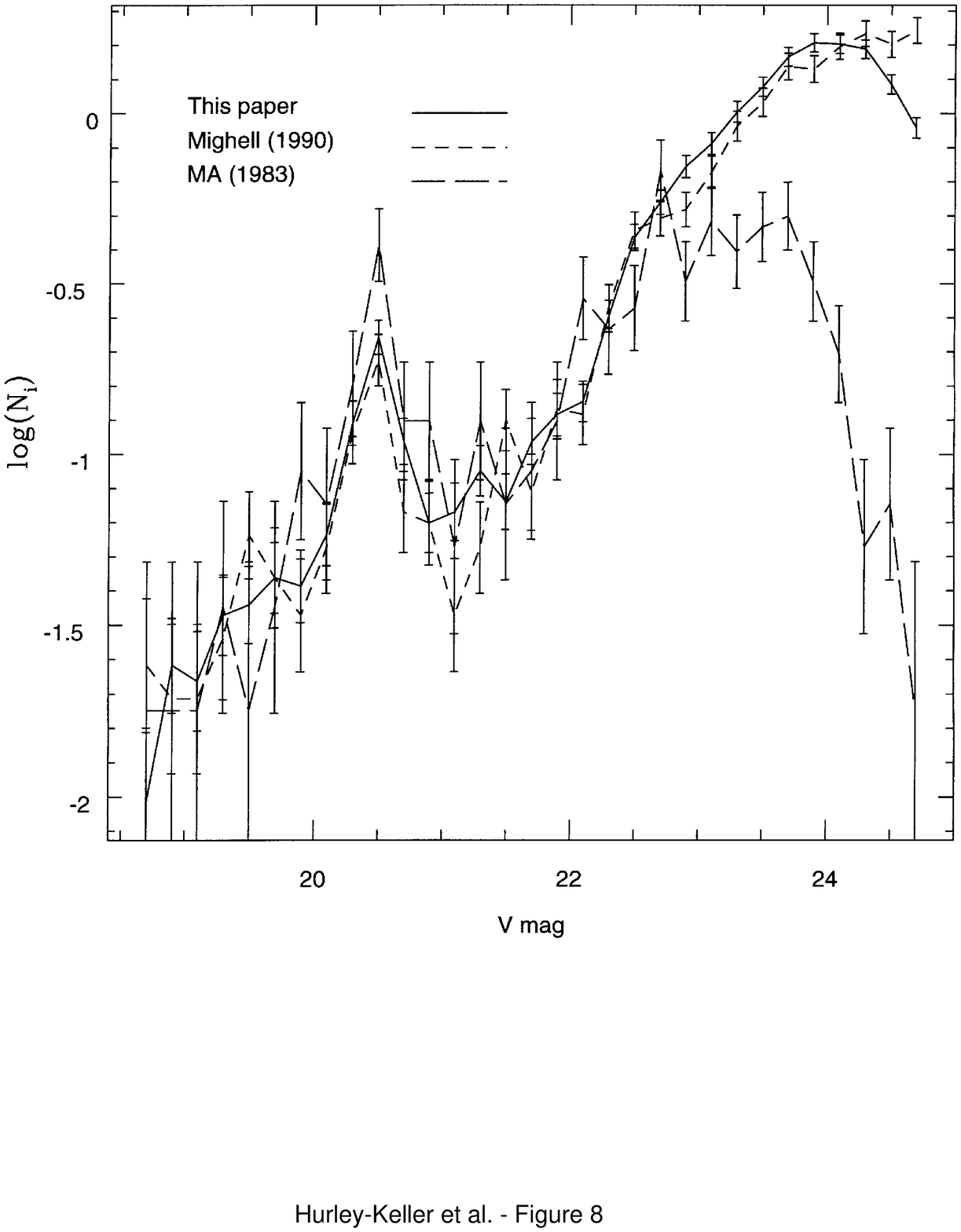}
\newpage
\plotone{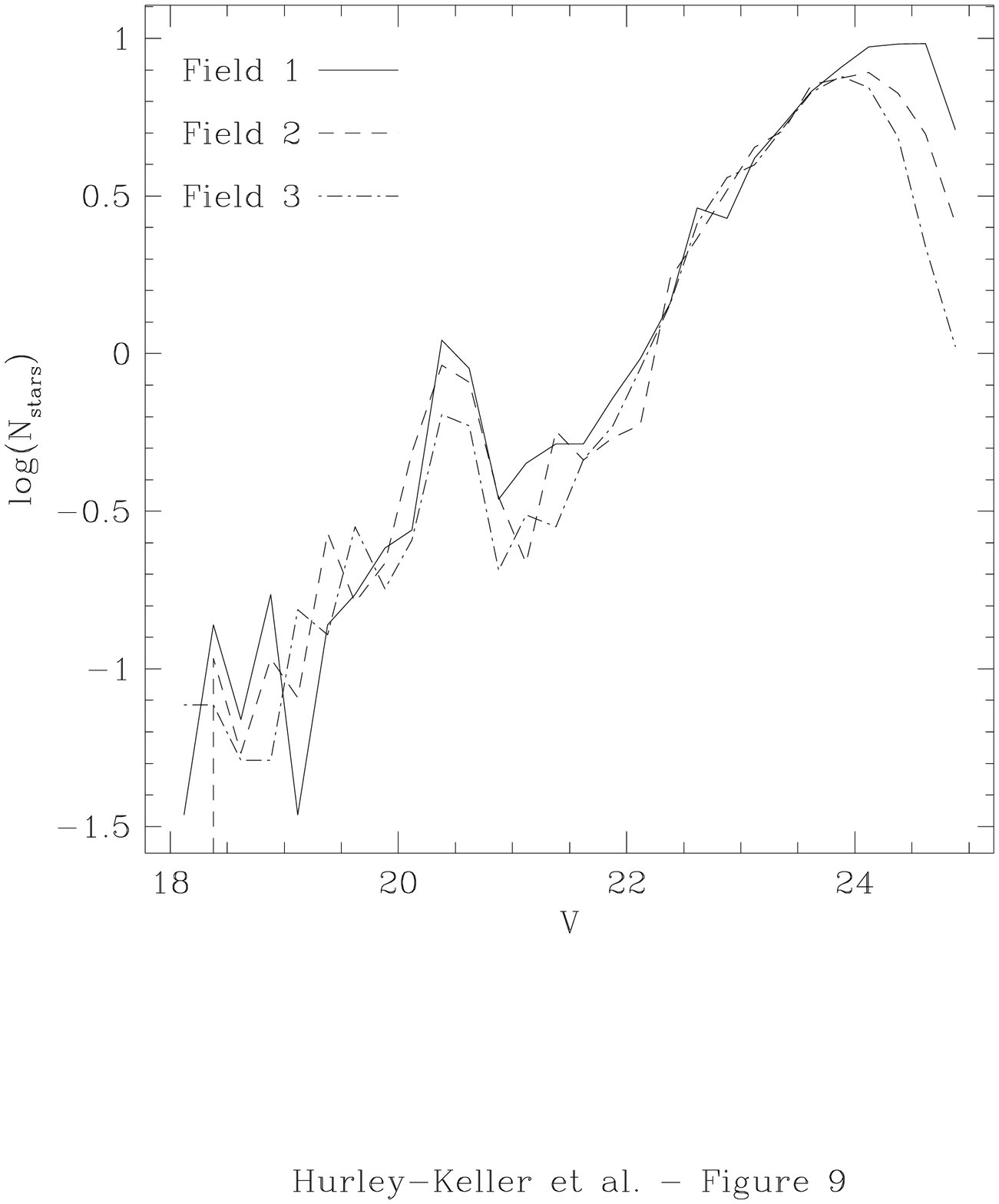}
\newpage
\plotone{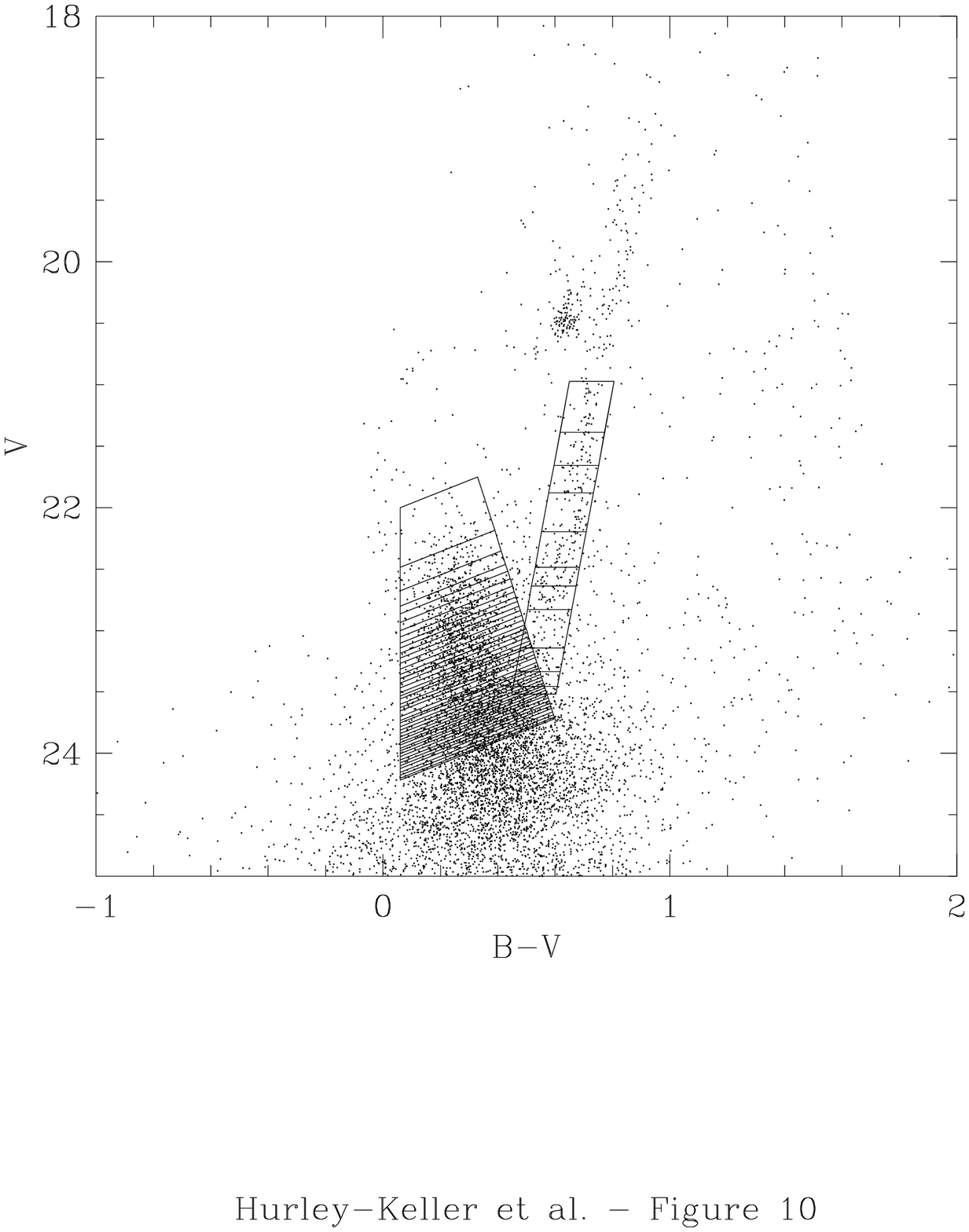}
\newpage
\plotone{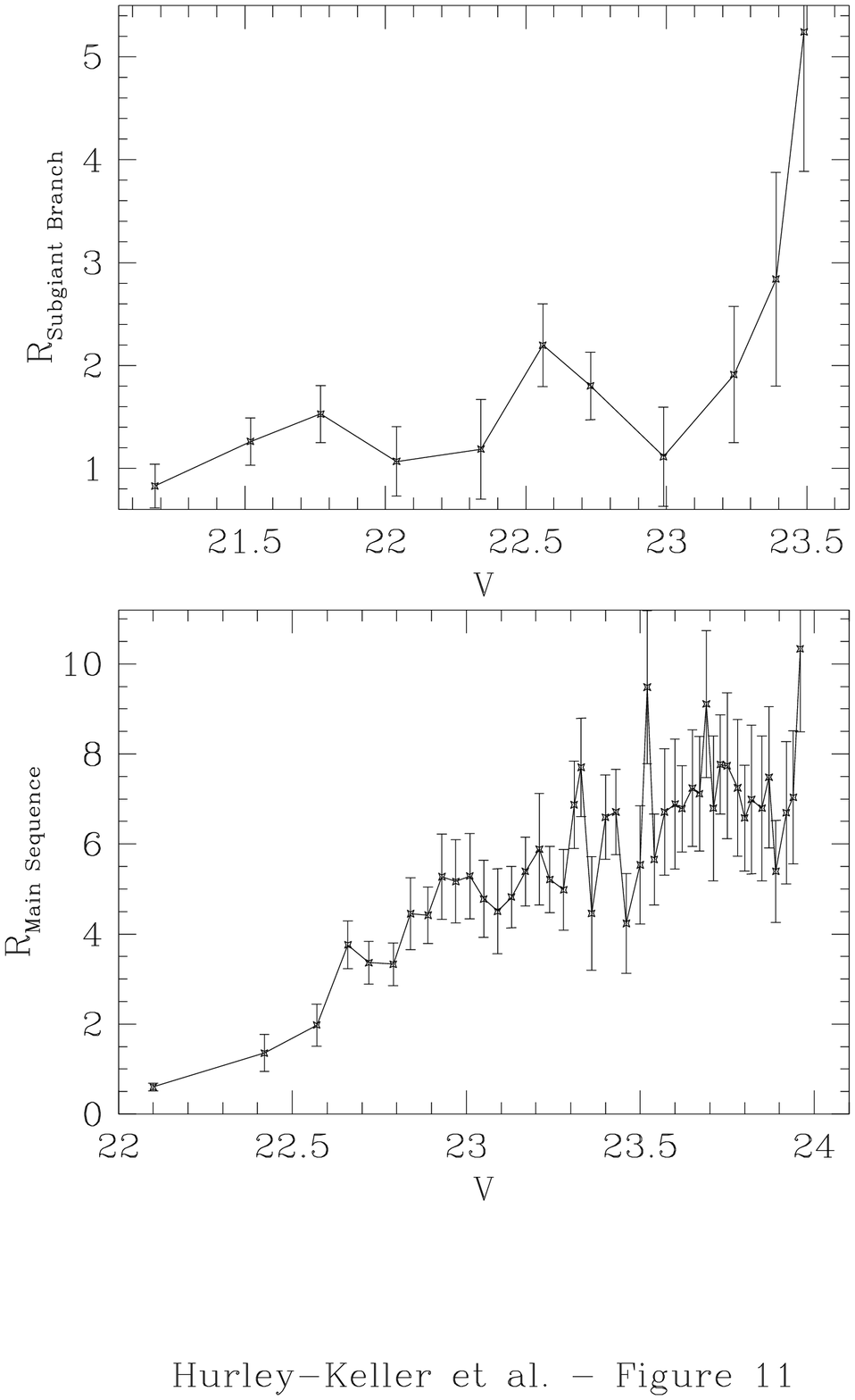}
\newpage
\plotone{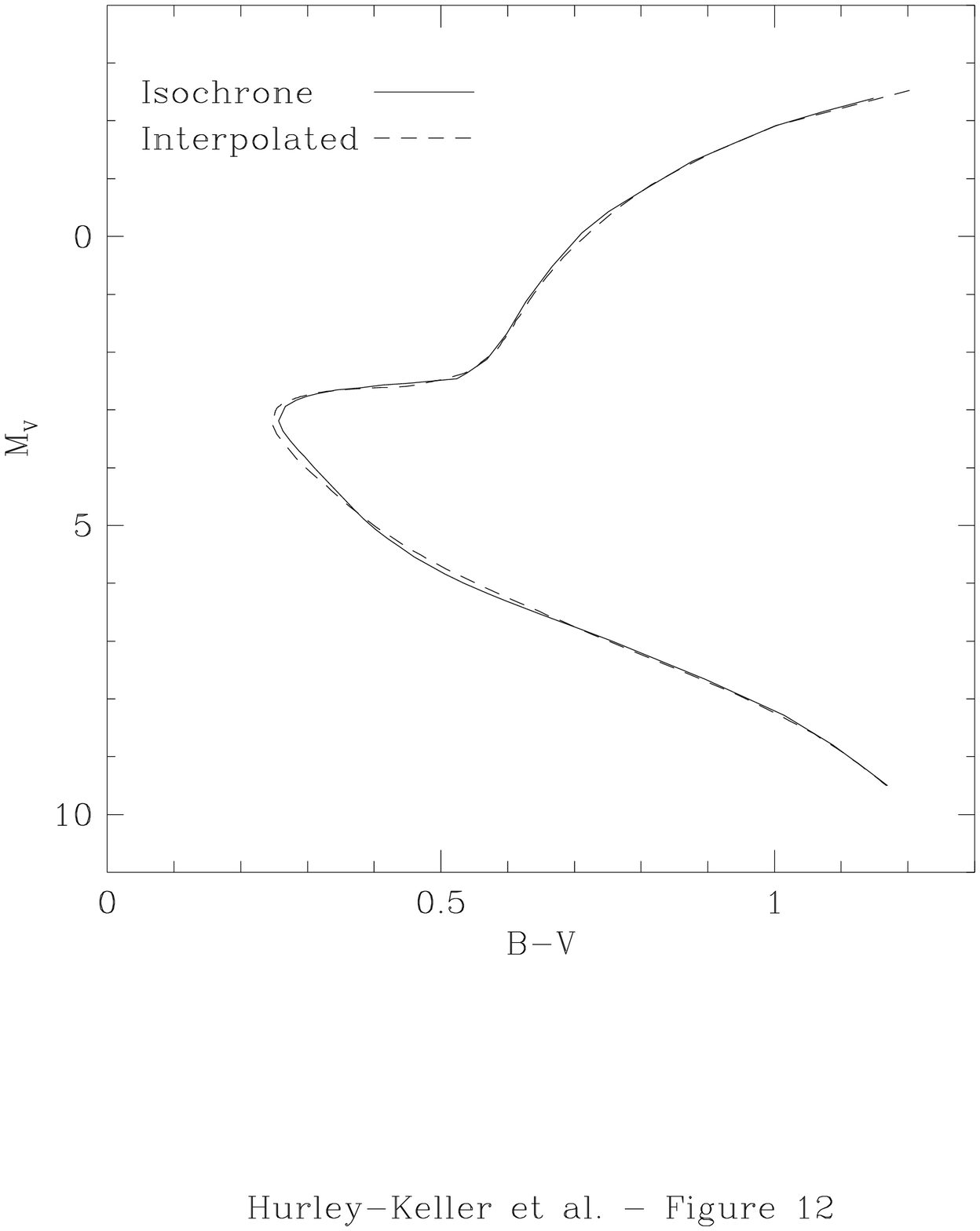}
\newpage
\plotone{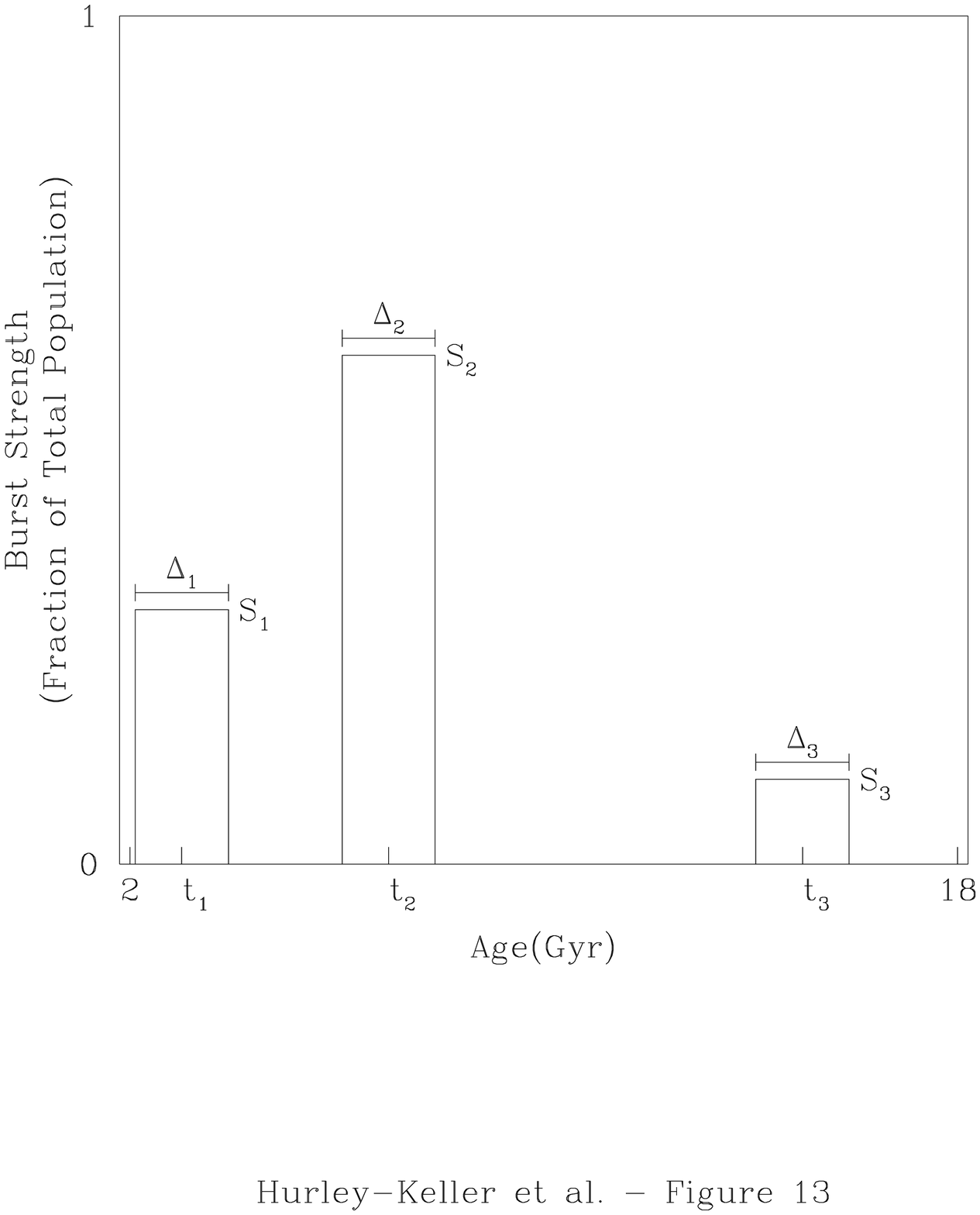}
\newpage
\plotone{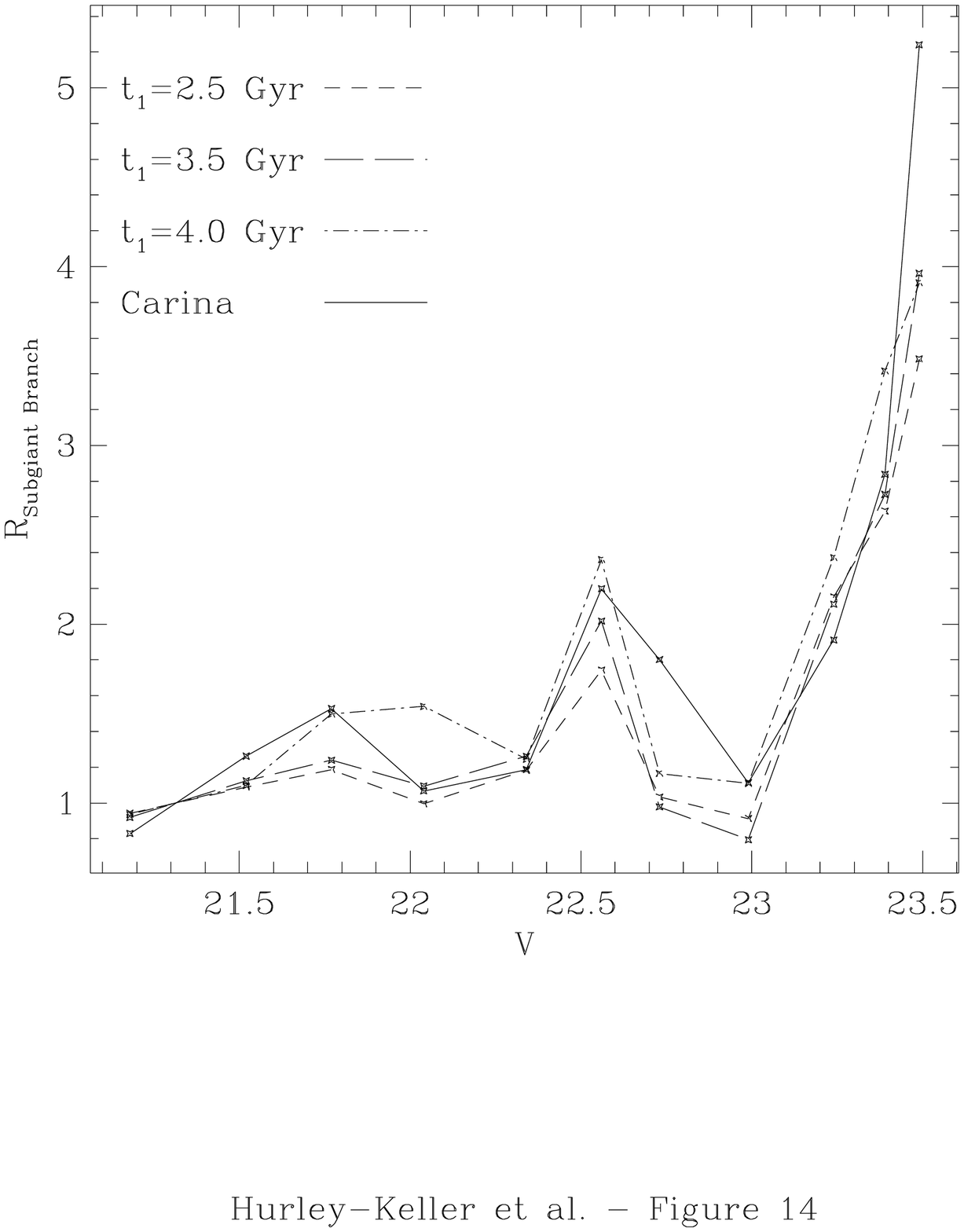}
\newpage
\plotone{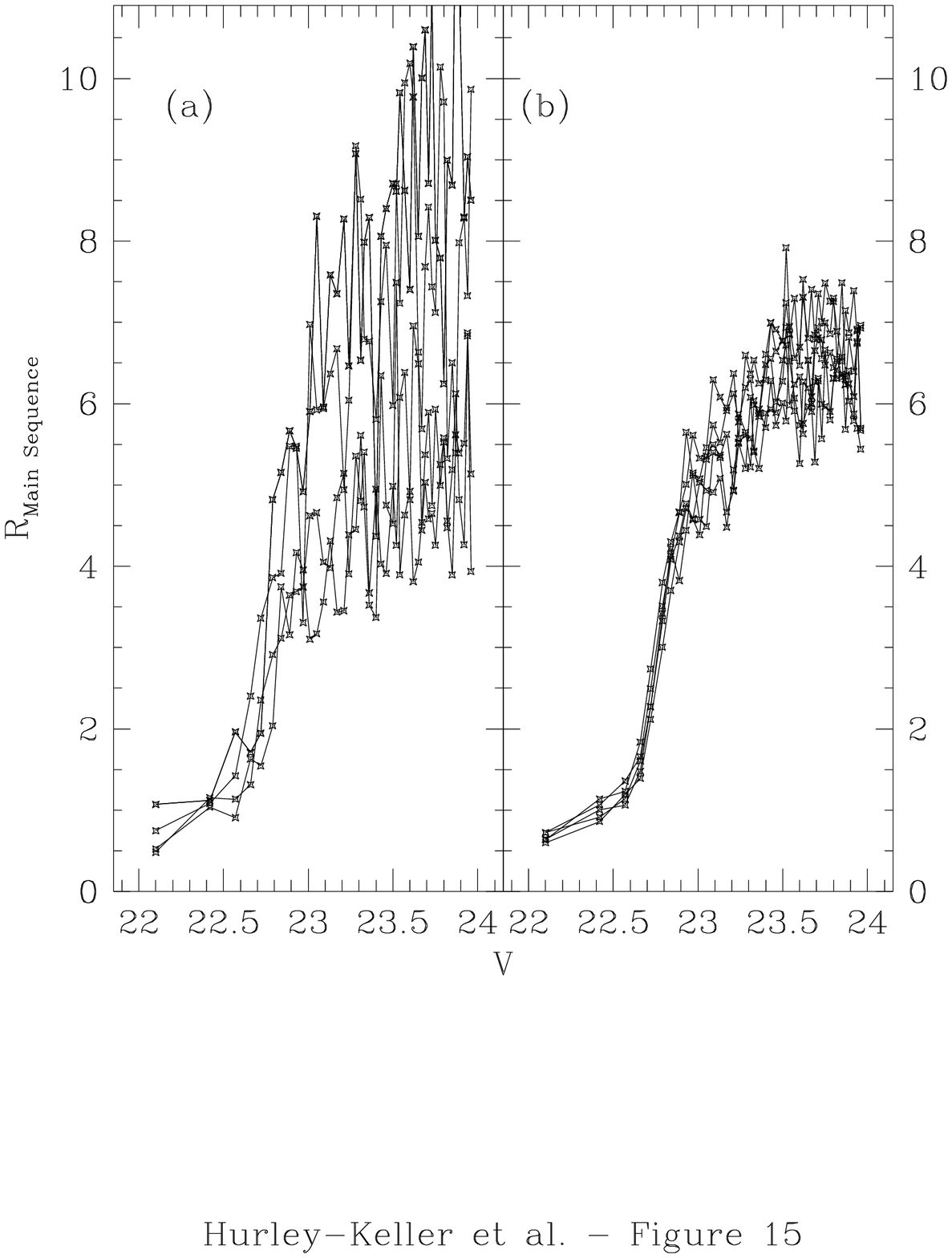}
\newpage
\plotone{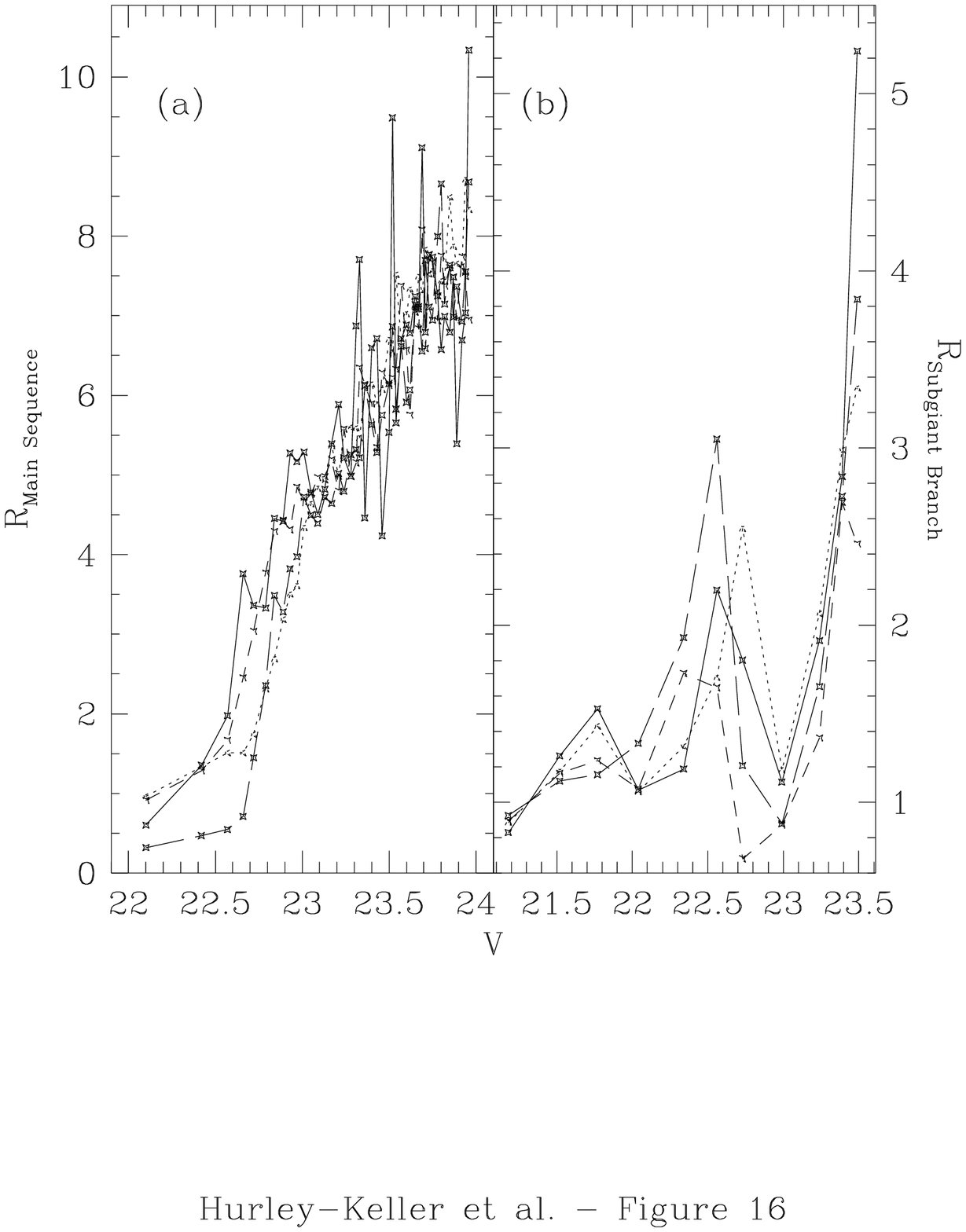}
\newpage
\plotone{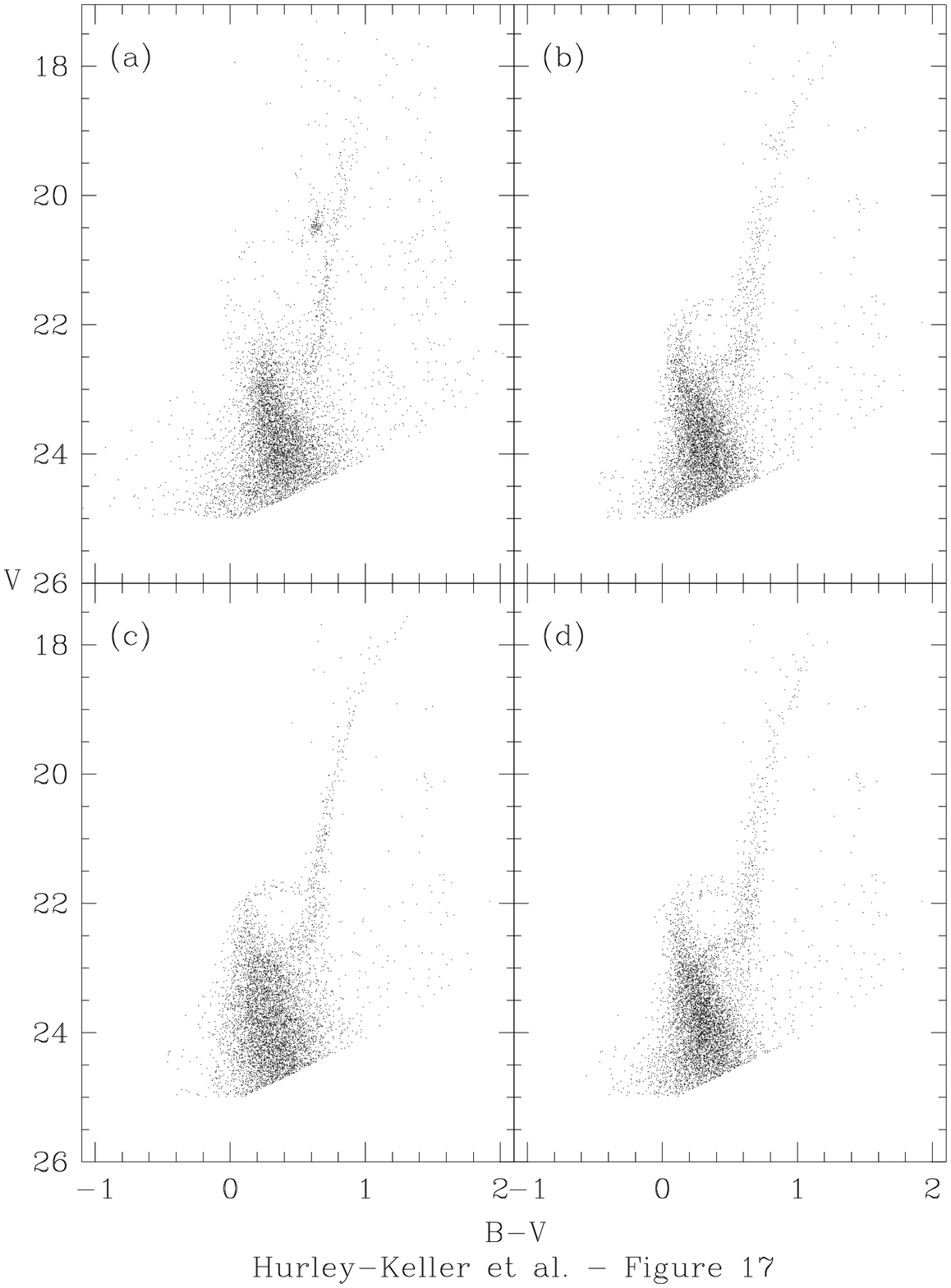}
\newpage
\plotone{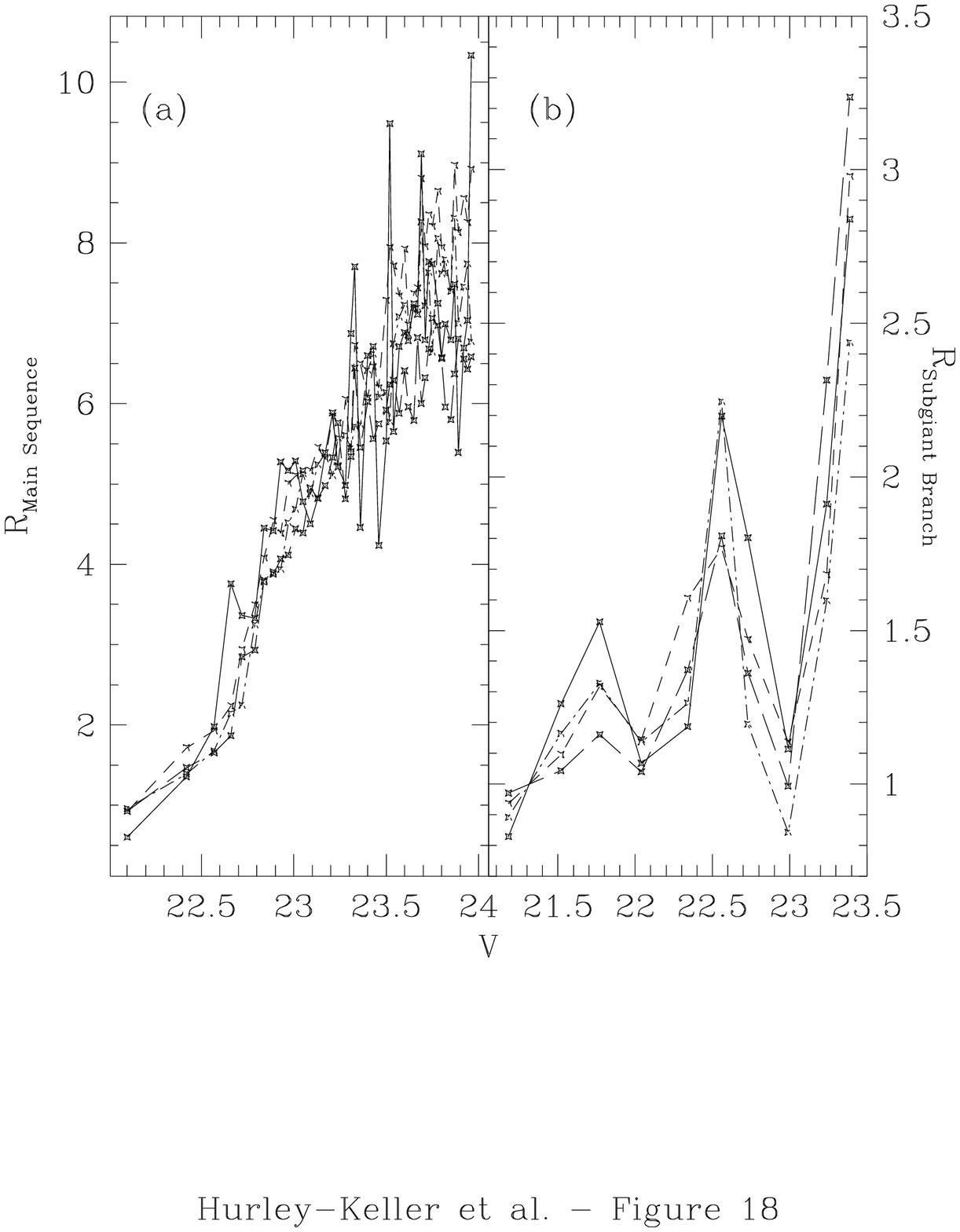}
\newpage
\plotone{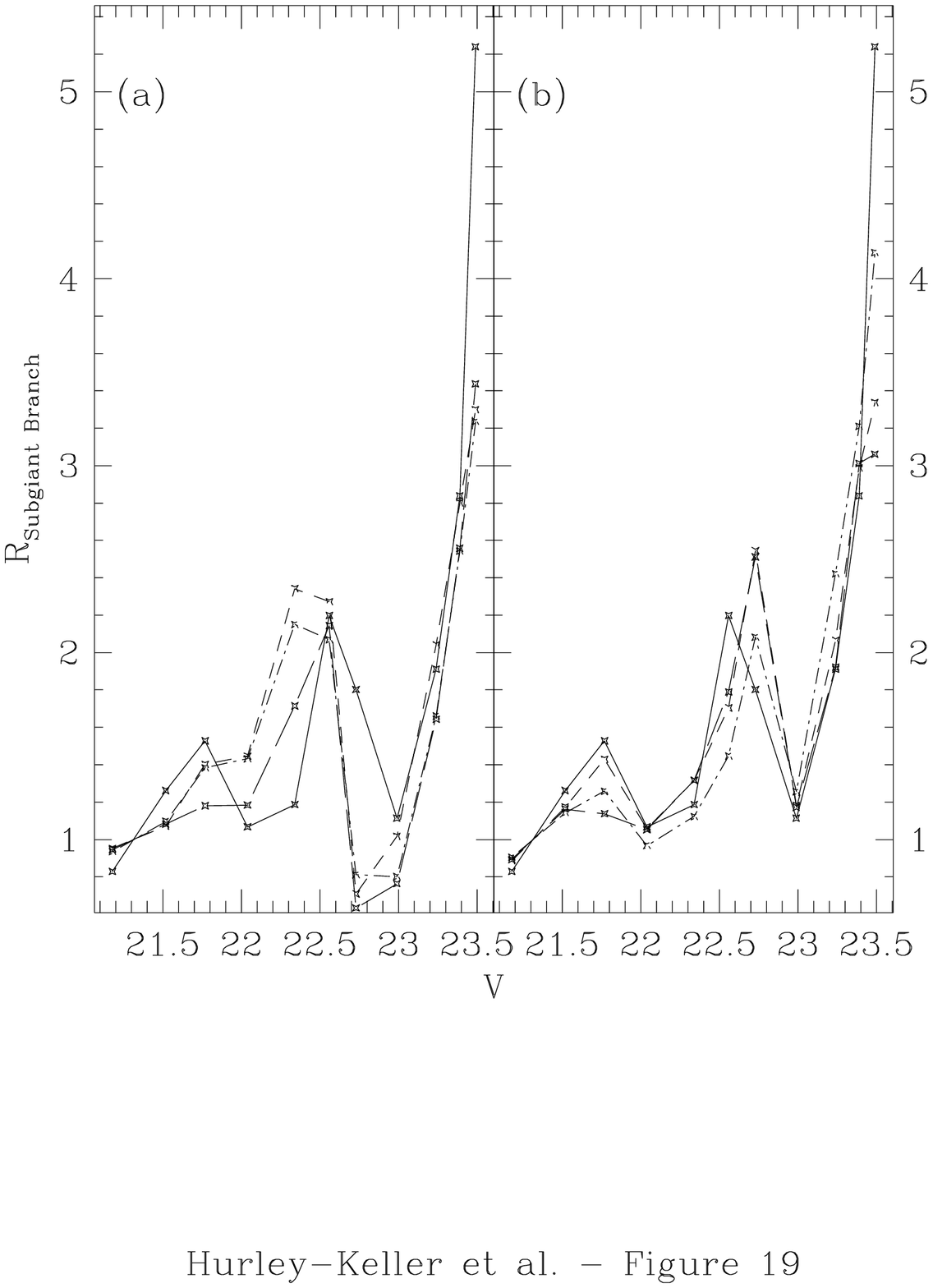}
\newpage
\plotone{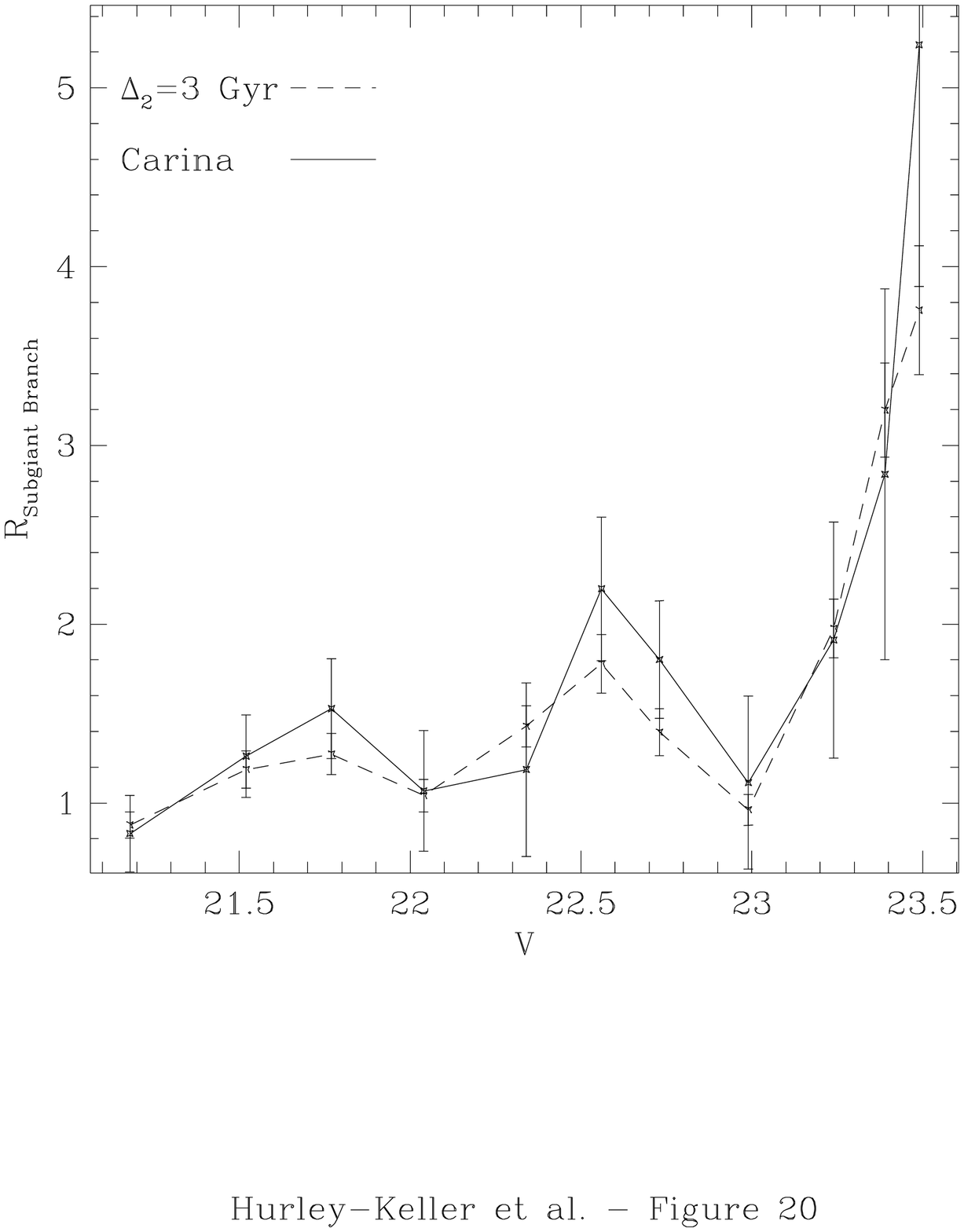}
\newpage

\begin{deluxetable}{cccccccc}
\tablecolumns{8}
\tablewidth{0pc}
\tablenum{1}
\tablecaption{Summary of Observations}
\tablehead{
\colhead{Field}		  & \colhead{$\alpha_{1950}$}	&
\colhead{$\delta_{1950}$} & \colhead{$N_{V}$\tablenotemark{a}}	&
\colhead{$N_{B}$\tablenotemark{a}} & 
\colhead{$t_{tot,V}$\tablenotemark{b}} &
\colhead{$t_{tot,B}$\tablenotemark{b}}	& 
\colhead{$N_{CMD}$\tablenotemark{c}}   \\
\colhead{} & \colhead{} & \colhead{} & \colhead{} &
\colhead{} & \colhead{(sec)} & \colhead{(sec)} & \colhead{} }
\startdata
1 & $06^{h}39^{m}53^{s}$ & $-50\arcdeg 56\arcmin 59\arcsec $ & 26 & 14 & 15721 & 6920 & 3051 \nl
2 & $06^{h}40^{m}07^{s}$ & $-50\arcdeg 52\arcmin 52\arcsec $ & 14 & 6 & 7000 & 3000 & 2233 \nl
3 & $06^{h}40^{m}42^{s}$ & $-50\arcdeg 52\arcmin 18\arcsec $ & 15 & 6 & 7800 & 3500 & 2025 \nl
Control ($1\arcdeg$ S) & $06^{h}40\arcmin $ & $-52\arcdeg$ & 4 & 4 & 2000 & 2000 & 440 \nl
\enddata
\tablenotetext{a}{$N_{i}$ refers to the total number of frames obtained in filter $i$}
\tablenotetext{b}{$t_{tot,i}$ refers to the total exposure time in filter $i$}
\tablenotetext{c}{$N_{CMD}$ refers to the total number of stars detected in the given field}
\end{deluxetable}

\begin{deluxetable}{crrrrrr}
\tablewidth{0pc}
\tablenum{2}
\tablecaption{Sample Photometry for Field 1}
\tablehead{
\colhead{Id}	& \colhead{X}	& \colhead{Y}	&
\colhead{V}	& \colhead{$\sigma_V$}		& 
\colhead{B-V}	& \colhead{$\sigma_{B-V}$}}

\startdata
    1 & 311.44 &   1.80 & 25.494 &  0.213 &  -0.143 &  0.298   \nl
    8 &  66.11 &   2.23 & 24.992 &  0.125 &   0.221 &  0.250	\nl   
    9 & 237.59 &   2.25 & 23.334 &  0.037 &   0.177 &  0.053	\nl   
   11 & 139.60 &   2.31 & 25.443 &  0.145 &  -0.043 &  0.245	\nl   
   10 & 196.25 &   2.31 & 23.715 &  0.037 &   0.238 &  0.057	\nl   
   12 &  85.34 &   2.41 & 25.289 &  0.172 &  -0.026 &  0.243	\nl   
   13 & 385.22 &   2.57 & 25.109 &  0.122 &   0.364 &  0.269	\nl   
   14 &  68.34 &   2.61 & 24.540 &  0.086 &   0.283 &  0.153	\nl   
   16 & 113.73 &   2.93 & 22.742 &  0.024 &   0.162 &  0.035	\nl   
   18 &  63.79 &   3.05 & 23.028 &  0.035 &   0.424 &  0.056	\nl   
   17 &  78.24 &   3.05 & 25.124 &  0.136 &   0.428 &  0.205	\nl   
   20 & 215.23 &   3.10 & 24.046 &  0.051 &   0.274 &  0.083	\nl   
   21 & 118.54 &   3.15 & 24.846 &  0.122 &   0.090 &  0.165	\nl   
   23 & 277.50 &   3.27 & 24.289 &  0.076 &   0.219 &  0.114	\nl   
   24 & 392.81 &   3.36 & 23.433 &  0.040 &   0.416 &  0.075	\nl   
   26 &  91.67 &   3.42 & 23.632 &  0.043 &   0.493 &  0.082	\nl   
   29 & 175.29 &   3.46 & 24.159 &  0.062 &   0.355 &  0.108	\nl   
   31 & 387.92 &   3.60 & 25.842 &  0.253 &  -0.164 &  0.402	\nl   
   36 & 381.73 &   3.87 & 25.108 &  0.114 &   0.445 &  0.256	\nl   
   39 & 283.09 &   3.97 & 24.173 &  0.065 &   0.203 &  0.102	\nl   
   38 & 318.18 &   3.97 & 25.188 &  0.129 &   0.330 &  0.261	\nl   
   40 & 126.18 &   4.06 & 24.097 &  0.066 &   0.686 &  0.136	\nl   
   43 & 218.48 &   4.10 & 25.117 &  0.118 &   0.107 &  0.189	\nl   
   44 & 111.47 &   4.13 & 25.301 &  0.186 &   0.072 &  0.309	\nl   
   45 &  86.80 &   4.25 & 24.993 &  0.116 &   0.058 &  0.177	\nl   
   46 &  58.36 &   4.42 & 24.009 &  0.053 &   0.441 &  0.124	\nl   
   49 & 380.40 &   4.55 & 25.597 &  0.171 &  -0.069 &  0.286	\nl   
   52 & 320.41 &   4.92 & 24.480 &  0.072 &   0.655 &  0.153	\nl   
   56 & 163.84 &   5.41 & 24.933 &  0.128 &  -0.001 &  0.170	\nl   
   57 & 309.76 &   5.49 & 24.102 &  0.049 &   0.272 &  0.090	\nl   
   58 & 257.79 &   5.53 & 22.833 &  0.028 &   0.216 &  0.040	\nl   
   59 & 191.60 &   5.57 & 24.933 &  0.091 &   0.565 &  0.213	\nl   
   60 & 263.86 &   5.61 & 24.726 &  0.082 &   0.272 &  0.145	\nl   
\enddata
\end{deluxetable}

\begin{deluxetable}{lcc}
\tablewidth{0pc}
\tablenum{3}
\tablecaption{Magnitude and Color Errors as a Function of Magnitude.}
\tablehead{
\colhead{$V$}		& \colhead{$\sigma_{v}$}		&
\colhead{$\sigma_{B-V}$}}

\startdata
17.80 & 0.02 & 0.03 \nl
18.80 & 0.01 & 0.02 \nl
19.80 & 0.01 & 0.02 \nl
20.80 & 0.01 & 0.02 \nl
21.80 & 0.02 & 0.04 \nl
22.80 & 0.04 & 0.08 \nl
23.80 & 0.10 & 0.18 \nl
24.80 & 0.16 & 0.20 \nl
25.80 & 0.23 & 0.25 \nl
\enddata
\end{deluxetable}

\begin{deluxetable}{lcccc}
\tablewidth{0pc}
\tablenum{4}
\tablecaption{Blue Straggler, Horizontal Branch, and Main-Sequence Turn-Off Stars in the Carina Fields}
\tablehead{
\colhead{}	& \colhead{BS\tablenotemark{a}}		& \colhead{BHB\tablenotemark{b}}		&
\colhead{Red Clump}	& \colhead{MSTO\tablenotemark{c}}}

\startdata
Field 1 & 19 & 2 & 31 & 87 \nl
Field 2 & 9 & 6 & 38 & 121 \nl
Field 3 & 15 & 2 & 23 & 133 \nl
\enddata
\tablenotetext{a}{BS refers to blue straggler candidates.}
\tablenotetext{b}{BHB refers to blue horizontal branch stars.}
\tablenotetext{c}{MSTO refers to the upper main-sequence turn-off stars.}
\end{deluxetable}

\begin{deluxetable}{cccccc}
\tablewidth{0pc}
\tablenum{5}
\tablecaption{Summary of Models: Ages and Durations of Episodes}
\tablehead{
\colhead{$t_1$}	&	\colhead{$t_2$}	&	\colhead{$t_3$}	&
\colhead{$\Delta_1$}	&	\colhead{$\Delta_2$}	&	\colhead{$\Delta_3$}}

\startdata
15 & 8 & 3 & 0.5 & 2 & 1 \nl
15 & 8 & 3 & 0.5 & 2 & 0.5 \nl
15 & 8 & 3 & 0.5 & 1 & 1 \nl
15 & 8 & 3 & 0.5 & 1 & 0.5 \nl
15 & 8 & 3 & 0.5 & 0.5 & 1 \nl
15 & 8 & 3 & 0.5 & 0.5 & 0.5 \nl
\hline
15 & 7 & 3 & 0.5 & 2 & 1 \nl
15 & 7 & 3 & 0.5 & 2 & 0.5 \nl
15 & 7 & 3 & 0.5 & 1 & 1 \nl
15 & 7 & 3 & 0.5 & 1 & 0.5 \nl
15 & 7 & 3 & 0.5 & 0.5 & 1 \nl
15 & 7 & 3 & 0.5 & 0.5 & 0.5 \nl
\hline
15 & 6.5 & 3 & 0.5 & 2 & 1 \nl
15 & 6.5 & 3 & 0.5 & 2 & 0.5 \nl
15 & 6.5 & 3 & 0.5 & 1 & 1 \nl
15 & 6.5 & 3 & 0.5 & 1 & 0.5 \nl
15 & 6.5 & 3 & 0.5 & 0.5 & 1 \nl
15 & 6.5 & 3 & 0.5 & 0.5 & 0.5 \nl
\enddata
\tablecomments{All values are in Gyr.}
\end{deluxetable}

\begin{deluxetable}{ccc}
\tablewidth{0pc}
\tablenum{6}
\tablecaption{Summary of Models: Strengths}
\tablehead{
\colhead{$S_1$}	&	\colhead{$S_2$}	&	\colhead{$S_3$}}

\startdata
0.1 & 0.7 & 0.2 \nl
0.1 & 0.6 & 0.3 \nl
0.1 & 0.5 & 0.4 \nl
\hline
0.2 & 0.7 & 0.1 \nl
0.2 & 0.6 & 0.2 \nl
0.2 & 0.5 & 0.3 \nl
0.2 & 0.4 & 0.4 \nl
\hline
0.3 & 0.6 & 0.1 \nl
0.3 & 0.5 & 0.2 \nl
0.3 & 0.4 & 0.3 \nl
\enddata
\tablecomments{$S_i$ is the fraction of the total population.}
\end{deluxetable}

\begin{deluxetable}{lccc}
\tablewidth{0pc}
\tablenum{7}
\tablecaption{Best Models}
\tablehead{
\colhead{Model}	& \colhead{$\chi_{MS}^2\%$\tablenotemark{a}}	& \colhead{$\chi_{SGB}^2\%$\tablenotemark{a}}	& \colhead{$\chi^2\%$\tablenotemark{a,b} }}

\startdata
$t_2=7.0; \Delta_2=2.0; \Delta_1=1.0 $ & 82 & 91 & 86.5 \nl
\hspace{10pt}$ S_1,S_2,S_3=.3, .5, .2$ & & & \nl
$t_2=7.0; \Delta_2=2.0; \Delta_1=0.5 $ & 84 & 88 & 86  \nl
\hspace{10pt}$S_1,S_2,S_3=.3, .5, .2$ & & & \nl
$t_2=7.0; \Delta_2=1.0; \Delta_1=1.0 $ & 84 & 87 & 85.5 \nl
\hspace{10pt}$S_1,S_2,S_3=.3, .5, .2$ & & & \nl
$t_2=7.0; \Delta_2=1.0; \Delta_1=0.5 $ & 80 & 80 & 80  \nl
\hspace{10pt}$S_1,S_2,S_3=.3, .5, .2$ & & & \nl
$t_2=7.0; \Delta_2=2.0; \Delta_1=1.0 $ & 75 & 87 & 81  \nl
\hspace{10pt}$S_1,S_2,S_3=.2, .6, .2$ & & & \nl
$t_2=7.0; \Delta_2=1.0; \Delta_1=0.5 $ & 72 & 90 & 81 \nl
\hspace{10pt}$S_1,S_2,S_3=.2, .6, .2$ & & & \nl
$t_2=7.0; \Delta_2=0.5; \Delta_1=1.0 $ & 60 & 79 \nl
\hspace{10pt}$S_1,S_2,S_3=.2, .6, .2$ & & & \nl
$t_2=6.5; \Delta_2=1.0; \Delta_1=0.5 $ & 58 & 66 \nl
\hspace{10pt}$S_1,S_2,S_3=.2, .5, .3$ & & & \nl
$t_2=7.0; \Delta_2=2.0; \Delta_1=1.0 $ & 57 & 84 \nl
\hspace{10pt}$S_1,S_2,S_3=.2, .7, .1$ & & & \nl
$t_2=6.5; \Delta_2=2.0; \Delta_1=0.5 $ & 54 & 61 \nl
\hspace{10pt}$S_1,S_2,S_3=.2, .5, .3$ & & & \nl
$t_2=7.0; \Delta_2=0.5; \Delta_1=1.0 $ & 54 & 67 \nl
\hspace{10pt}$S_1,S_2,S_3=.3, .5, .2$ & & & \nl
$t_2=7.0; \Delta_2=0.5; \Delta_1=0.5 $ & 48 & 80 \nl
\hspace{10pt} $S_1,S_2,S_3=.3, .5, .2$ & & & \nl
\enddata
\tablenotetext{a}{$\chi_{MS}^2 \%$, $\chi_{SGB}^2 \%$, and average $\chi^2 \%$ are
actually the $\chi^2$ probabilities for the models computed $\chi^2$ value.}
\tablenotetext{b}{The average $\chi^2 \%$ is shown only for those models which
passed the selection criteria.}
\end{deluxetable}

\begin{deluxetable}{lccc}
\tablewidth{0pc}
\tablenum{8}
\tablecaption{Summary of Best Model Properties}
\tablehead{
\colhead{SFH Parameters}	& \colhead{$t_2=6.5$}	& \colhead{$t_2=7$}	& \colhead{$t_2=8$}}

\startdata
$\Delta_2=0.5$ & 0 & 3 & 0 \nl
$\Delta_2=1.0$ & 1 & 3 & 0 \nl
$\Delta_2=2.0$ & 1 & 4 & 0 \nl
\hline
$S_3=0.1$ & 0 & 1 & 0 \nl
$S_3=0.2$ & 0 & 9 & 0 \nl
$S_3=0.3$ & 2 & 0 & 0 \nl
\hline
$S_2=S_1$ & 0 & 0 & 0 \nl
$S_2>S_1$ & 2 & 10 & 0 \nl
\hline
$\Delta_1=0.5$ & 2 & 4 & 0 \nl
$\Delta_1=1.0$ & 0 & 6 & 0 \nl
\enddata
\end{deluxetable}

\end{document}